\documentclass[12pt,letterpaper]{article}
\usepackage[T1]{fontenc}
\usepackage[utf8]{inputenc}
\usepackage{lmodern,microtype}
\usepackage[margin=1in]{geometry}
\usepackage{amsmath,amssymb,amsthm,mathtools}
\usepackage{booktabs,array,tikz,placeins}
\usepackage{enumitem,setspace}
\usepackage[authoryear,round]{natbib}
\bibpunct{(}{)}{;}{a}{}{,}
\usepackage{xurl}
\usepackage[hidelinks]{hyperref}
\hypersetup{
  pdftitle={Verification and the Risk of a Wrong Cascade},
  pdfauthor={Darina Cheredina and Georgy Lukyanov}
}

\setlist{nosep,leftmargin=*}
\numberwithin{equation}{section}
\newtheorem{proposition}{Proposition}[section]
\newtheorem{theorem}[proposition]{Theorem}
\newtheorem{lemma}[proposition]{Lemma}
\newtheorem{corollary}[proposition]{Corollary}
\theoremstyle{definition}
\newtheorem{definition}[proposition]{Definition}

\theoremstyle{remark}
\newtheorem{remark}[proposition]{Remark}
\newtheorem{example}[proposition]{Example}

\title{Verification and the Risk of a Wrong Cascade\thanks{We are grateful to Evgeniy Andreev, Emiliano Catonini, Markus Gebauer, Ella Khromova, Steven Kivinen, and Alexey Verenikin for helpful comments and discussions, and to participants in the ICEF research seminar for valuable feedback. Financial support from the French National Research Agency (ANR) under grant ANR-17-EURE-0010 (Investissements d'Avenir program) is gratefully acknowledged. All remaining errors are our own.}}
\author{Darina Cheredina\thanks{HSE University, International College of Economics
and Finance, 1 Pokrovsky Bulvar, building T, Moscow, 109028, Russia.}
\and Georgy Lukyanov\thanks{Toulouse School of Economics, 1 Esplanade de
l'Universit\'e, Toulouse, 31080, France. Corresponding author. Email:
\href{mailto:georgy.lukyanov@tse-fr.eu}{georgy.lukyanov@tse-fr.eu}.}}
\date{}

\begin{document}
\maketitle
\begin{abstract}
We study social learning when agents can pay for public evidence after seeing a
private signal. A successful test establishes state one and earns its investigator
a bounty, even when the information cannot improve her own action. With a common
positive cost and an imperfect test, the public reaches proof or a permanent
cascade in finite expected time. In one specification with a symmetric prior,
verification eliminates wrong cascades in state zero and makes them more likely in
state one for every private-signal precision. A favorable signal that would have
started a correct cascade can instead induce a test. If it fails, later agents may
abandon the correct action. For a family of signal and test precisions we derive
the three terminal probabilities in closed form. A larger bounty can increase both
the probability of proof and the probability of a wrong cascade in state one.

\end{abstract}
\medskip
\noindent\textit{Keywords:} Information cascades, Social learning, Costly verification, Information acquisition, Hard evidence, Bayesian learning
\par\noindent\textit{JEL classification:} D83; D82; C73
\medskip

\section{Introduction}\label{sec:introduction}

Suppose that several firms use machines of the same design. Each firm observes how
its own machine performs and decides, in turn, whether to replace it. A test can
sometimes establish that the design is faulty. For a firm that intends to replace
its machine whatever the test finds, the information has no value for its own
decision. It may nevertheless run the test if the manufacturer offers a bounty for
documenting a fault. Consider now the next firm. A replacement alone might tell it
very little, especially if its predecessors have all replaced their machines. The
decision to test may tell it something about what its predecessor saw. The outcome
of that test gives it a further reason to revise its belief.

In our model each agent takes one of two actions. She would like her action to
match the state, and she observes both the public history and a private signal
before deciding whether to investigate. An investigation succeeds with probability
$p$ in state one and always fails in state zero. The agent who succeeds earns a
fixed bounty. Everyone sees whether she investigated, what she found, and which
action she took. If both signal types decline to test and choose the same action,
the next agent faces exactly the same decision. An information cascade has begun.
We ask how access to verification changes the probability that the action taken in
such a cascade is wrong.

Consider the first agent when the two states are equally likely and her signal is
correct with probability $3/4$. If she cannot investigate, a favorable signal leads
her to take action one. The next agent then takes the same action even after an
unfavorable signal. Suppose instead that verification is available at the cost and
bounty of Example~\ref{ex:disconnected-region}. The favorable type investigates.
Her decision tells the public what her action alone would have told it. A failed
test then reduces the public belief in state one from $3/4$ to $3/5$. At this
belief, two consecutive unfavorable signals suffice for later agents to settle on
action zero. If the true state is one, an opportunity to produce proof has opened
up a way of abandoning the correct action.

How often does this happen? With verification the probability of a wrong lower
cascade in state one rises from $1/13$ to about $0.1415$. In state zero the
probability of a wrong upper cascade falls from $4/13$ to zero. At this precision,
whenever an upper cascade could otherwise begin, the favorable type expects a
bounty large enough to justify testing. Her test always fails in state zero.
Proposition~\ref{prop:submitted-error-comparison} establishes both comparisons for
every private-signal precision strictly between one half and one, keeping the
prior, cost, bounty, and test technology fixed. With equal prior weights
verification improves the accuracy of the eventual action.\footnote{Terminal
accuracy gives the same weight to two histories that end with the same action in
the same state, even if one takes longer and involves more costly investigations.}

The effect in state one can go the other way. To calculate it we first establish
that, with a common positive cost and an imperfect test, the public reaches proof
or a cascade in finite expected time. We can then compare the probabilities of the
same cascade history in the two states using the public odds at which it begins.
When a favorable signal and a failed test exactly offset each other in odds, this
argument gives closed forms for every cost and bounty. At $q=2/3$ and the lower
cost $c=1/5$ of Example~\ref{ex:two-motives}, the state-one error is $1/7$ without
verification and $1/9$ with a bounty of $1/4$. Raising the bounty to $1/2$ brings
the error up to $1/8$. More agents now produce proof, but some histories that used
to end in a correct optimistic cascade end in a mistaken pessimistic one.

Whether agents continue to investigate an optimistic consensus depends on what the
sponsor rewards. An agent may test to choose a better action or simply to collect
the bounty. If the payment falls as proof becomes more widely anticipated, even the
favorable type may decline at a belief where she would have tested for a fixed
payment. There is a second possibility. The agent may be willing to test at
intermediate and at very high private beliefs, with a gap in between. We compare
two policies under which the testing decisions differ only inside this gap and
calculate the change in each state's error.

Section~\ref{sec:literature} discusses related work. We introduce the model in
Section~\ref{sec:model} and derive the private incentive to investigate in
Section~\ref{sec:motives}. Section~\ref{sec:dynamics} follows public beliefs until
learning stops and compares the terminal outcomes with and without verification.
Section~\ref{sec:robustness} considers false positives and privately observed
costs. \hyperref[app:proofs]{Appendix~A} gives the longer proofs. The histories and calculations behind
the tables are in \hyperref[app:calculations]{Online Appendix~B}.

\section{Related literature}\label{sec:literature}

In the models of \citet{Banerjee1992} and \citet{BikhchandaniHirshleiferWelch1992},
once the same action is optimal after either private signal, observing another
agent take it reveals nothing. Whether private beliefs have bounded support matters
for long-run learning
\citep{SmithSorensen2000,BikhchandaniHirshleiferTamuzWelch2024}. Our question is
close to the one asked by \citet[Proposition~5]{HendricksSorensenWiseman2012}. They
calculate how often consumers give up searching when the product is in fact of high
quality. Their restriction on the signal distribution prevents a jump across the
no-search boundary. In our model a single failed investigation can take public
beliefs past a cascade boundary. To calculate the probability of that cascade, we
need the odds where the public first enters it.

Several papers allow agents to choose how much information to acquire.
\citet{BurguetVives2000} study the limits on learning imposed by a positive
marginal cost of private precision. \citet{MuellerFrankPai2016} relate long-run
learning to the lower support of private search costs, and \citet{Ali2018} studies
herding when agents choose whether to acquire information. In the companion paper
by \citet{IvanikLukyanov2025}, the choice of experiment is based on public history.
Their Theorem~5.1 bounds the expected number of purchases when each informative
purchase has a minimum fee. In their binary example the public belief at purchase
dates follows a nearest-neighbor walk in log odds, and the terminal-belief errors
are symmetric (Theorem~5.5).

Here the agent who decides whether to buy a test has already received a signal. Her
willingness to pay can itself reveal that signal. Moreover, evidence can establish
only one of the states, and a bounty may cover the cost even when the information
has no value for the agent's action. We therefore need to show when these purchases
cease and to distinguish, within state one, discovery from a correct cascade in
which nobody has produced proof.

An unsuccessful experiment can also lead to abandonment in
\citet{KellerRadyCripps2005}. In their model a breakthrough on the risky arm
establishes that the state is good. As agents experiment without a breakthrough,
their belief in the good state falls. These agents are long-lived and can wait for
others to experiment. Each of our agents acts only once and decides whether to test
after seeing her private signal. Conditional on seeing her start the test, a
failure is always adverse news. But the attempt itself may have disclosed a
favorable signal. To know what the next agent has learned, one must ask whether the
action alone would have disclosed it as well.

The form of the transfer matters for what the public learns. In one mechanism of
\citet{SmithSorensenTian2021} an agent is paid according to whether her successor
follows her action. This links her incentives to the information her decision
supplies to others. \citet{FrazierKempeKleinbergKleinberg2014} study the payments
needed to induce exploration by myopic agents and the performance that exploration
can deliver. In the contests of \citet{HalacKartikLiu2017}, the division of the
prize and the disclosure of success both affect the incentive to experiment. Our
sponsor promises a payment to the investigator who produces a verifiable finding,
and disclosure is automatic. We take the payment rule as given and ask what happens
to the two errors if a discovery earns less when the public already expects it.

\section{Model}\label{sec:model}

Time is discrete. At each date $t$ a new agent chooses between actions zero and
one. The state $\theta\in\{0,1\}$ is common to all dates and has prior probability
$\pi=\Pr(\theta=1)\in(0,1)$. Before acting, agent $t$ observes the public history
$H_t$ and receives a private signal $s_t\in\{0,1\}$. Signals are independent across
dates conditional on $\theta$ and have precision $q$ given by
\begin{equation}\label{eq:signal-technology}
 \Pr(s_t=1\mid\theta=1)=\Pr(s_t=0\mid\theta=0)=q,\qquad 1/2<q<1.
\end{equation}
Let $\mu_t=\Pr(\theta=1\mid H_t)$ denote the belief of an observer who sees the
public history and $x_t$ the belief of agent $t$ once she also knows her signal.
Given an interior public belief $\mu$, Bayes' rule gives the two possible private
beliefs
\begin{equation}\label{eq:private-posteriors}
 x^1(\mu)=\frac{\mu q}{\mu q+(1-\mu)(1-q)},\qquad
 x^0(\mu)=\frac{\mu(1-q)}{\mu(1-q)+(1-\mu)q}.
\end{equation}

The agent can now pay $c>0$ to investigate. Her choice is $i_t=1$ if she runs the
test and $i_t=0$ if she abstains. Conditional on investigation the outcome $e_t$ is
public and has distribution
\begin{equation}\label{eq:verification-technology}
 \begin{aligned}
 \Pr(e_t=1\mid\theta=1,i_t=1)&=p,\\
 \Pr(e_t=1\mid\theta=0,i_t=1)&=0,
 \end{aligned}\qquad 0<p\leq1 .
\end{equation}
Conditional on the state, a test draw is independent of all other draws and of the
private signals. We call $e_t=1$ a success, since it proves state~1. A failed
investigation produces $e_t=0$. We record $e_t=\varnothing$ when the agent
abstains. A sponsor commits in advance to pay $V\geq0$ to an agent who produces
proof. In Section~\ref{subsec:reward-and-gap} we also allow this payment to depend
on the public belief just before the agent decides whether to test.

Having observed the outcome, if there is one, the agent chooses her action
$a_t\in\{0,1\}$. Her payoff is
\begin{equation}\label{eq:period-payoff}
 u_t=\mathbf1\{a_t=\theta\}-ci_t+V\mathbf1\{e_t=1\}.
\end{equation}
The agents who follow her observe $(i_t,e_t,a_t)$ in order. In particular, an
investigator cannot withhold a failure. Thus $H_t$ consists of the complete records
of agents $1,\ldots,t-1$. Once an outcome establishes the state the process ends.
With a perfect test ($p=1$), either outcome does so.

An agent who abstains chooses $a^\star(x)=\mathbf1\{x\geq1/2\}$ at private belief
$x$. An agent who obtains proof chooses one. If the test fails, her posterior
becomes
\begin{equation}\label{eq:failure-map}
 \Phi_p(x)=\frac{x(1-p)}{1-px}.
\end{equation}
whenever failure is feasible. For an interior $x$ this posterior is strictly below
$x$. She still takes action one after a failure exactly when
\begin{equation}\label{eq:post-failure-action-threshold}
 x\geq\bar x_p,\qquad \bar x_p=\frac{1}{2-p}.
\end{equation}
At $(p,x)=(1,1)$ failure has zero probability and only the action after success is
relevant.

We use perfect Bayesian equilibrium. Each agent maximizes her own period payoff,
and beliefs follow Bayes' rule at every feasible record. When indifferent between
the actions she takes action~1. When indifferent about investigating she abstains.
These conventions matter for the beliefs at which a cascade begins. Off the
equilibrium path, beliefs must respect conclusive evidence. Since an agent has no
stake in later actions, we can first solve her decision for a given private belief
and then ask what later agents infer from it.

\section{Who investigates?}\label{sec:motives}

Consider an agent with private belief $x$. If she abstains, her best action gives
her $A(x)=\max\{x,1-x\}$. Investigation gives her $xp(1+V)+(1-xp)A(\Phi_p(x))-c$.
The second term is zero if failure is impossible. Let $G(x)$ denote the difference
between these two payoffs, taking the payoff from investigation first. She tests
exactly when $G(x)>0$.

\begin{proposition}\label{prop:gain-decomposition}
The net gain from testing is
\begin{equation}\label{eq:gain-decomposition}
  G(x)=D_p(x)+xpV-c ,
\end{equation}
where the improvement in the agent's expected action payoff is
\begin{equation}\label{eq:diagnostic-value}
 D_p(x)=
 \begin{cases}
 px, &0\leq x\leq 1/2,\\
 1-(2-p)x, &1/2\leq x\leq \bar x_p,\\
 0, &\bar x_p\leq x\leq1 .
 \end{cases}
\end{equation}
\end{proposition}
\begin{proof}
Take first an agent with $x<1/2$. Without a test she chooses zero. Success allows
her to correct a mistake in state~1, an event of probability $xp$. If
$1/2\leq x<\bar x_p$, the agent would already choose one without a test. By acting
on its outcome she raises her probability of being correct from $x$ to $1-x+xp$.
Finally, at $x\geq\bar x_p$ she chooses one after every feasible outcome. These
expressions agree at either boundary. In all three cases her expected payment is
$xpV$.
\end{proof}

\begin{lemma}\label{lem:diagnostic-shape}
Diagnostic value is positive on $(0,\bar x_p)$, rises to its unique
maximum $D_p(1/2)=p/2$, and declines to zero at $\bar x_p$. At any
fixed posterior it is weakly increasing in $p$.
\end{lemma}
\begin{proof}
On the lower half of the belief interval the diagnostic gain is $px$. On the upper
half it is $\max\{1-(2-p)x,0\}$. The result follows on inspecting these two
expressions.
\end{proof}

Consider an investigator with $x\geq\bar x_p$. She expects to take the same action
whatever the test finds, so her whole return from testing is the expected bounty.
She tests if $xpV>c$. An agent closer to $1/2$ can also gain from changing her
action. To see how the two motives interact, write
\begin{equation}\label{eq:kappa-definition}
 \kappa=p(1+V),\qquad W=pV,\qquad
 \mathcal X_I(p,c,V)=\{x\in[0,1]:G(x)>0\}.
\end{equation}
It is enough to compare the cost with the gross return at the three beliefs $1/2$,
$\bar x_p$, and one. In this order the returns are
\begin{equation}\label{eq:critical-costs}
 c_M=\kappa/2,\qquad c_B=W/(2-p),\qquad c_T=W .
\end{equation}

\begin{theorem}\label{thm:investigation-geometry}
The testing set $\mathcal X_I$ consists of at most two intervals. A
component containing $1/2$ and one reaching belief one can be separated
only if $\kappa<2$. In that case their possible boundaries are
\begin{equation}\label{eq:declining-region-roots}
 \ell=c/\kappa,\qquad u=(1-c)/(2-\kappa),\qquad h=c/W
\end{equation}
with $h$ used when $W>0$. A separated pair is $(\ell,u)\cup(h,1]$,
including $u=h$ if a single indifferent posterior separates it. If
$\kappa\geq2$, any nonempty testing set is a single upper interval.
\end{theorem}
\begin{proof}
Starting at $G(0)=-c$, the gain has successive slopes $\kappa$, $\kappa-2$, and
$W$. A negative middle slope allows the gain to turn down and then up again.
Solving for the zero on each piece gives the cutoffs in the statement. With a
nonnegative middle slope the gain is weakly increasing over the entire interval.
The cost ranges for each case, including a flat middle piece, are given in
\ref{app:gain-boundaries}.
\end{proof}

\begin{corollary}\label{cor:investigation-components}
The testing set is nonempty exactly when $c<\max\{c_M,c_T\}$.
It has two components exactly when
\begin{equation}\label{eq:disconnected-condition}
 \kappa<2,\qquad c_B\leq c<\min\{c_M,c_T\}.
\end{equation}
\end{corollary}
\begin{proof}
The highest gross return occurs at the midpoint or at one. For there to be two
components, both returns must strictly exceed the cost, and the return at the
minimum between them must be at most the cost. This gives
\eqref{eq:disconnected-condition}, including the case in which a single indifferent
type separates the two intervals.
\end{proof}

\begin{example}\label{ex:disconnected-region}
Take $p=1/2$, $V=2$, and $c=7/10$. The rule is
\begin{equation}\label{eq:disconnected-example-region}
 \mathcal X_I=(7/15,3/5)\ \cup\ (7/10,1].
\end{equation}
Both components require the bounty because the largest diagnostic gain
is $1/4$, below the cost. In the first component a test can change the
action. Throughout the second even failure leaves action one optimal, so the
entire gross return is the expected payment.
Section~\ref{sec:dynamics} starts from this example.
\end{example}

\begin{example}\label{ex:two-motives}
Keep $p=1/2$, lower the cost to $1/5$, and take $q=2/3$ and
$\pi=1/2$. The private testing sets under three fixed bounties are
\[
 \begin{array}{c|l}
 V & \mathcal X_I\\ \hline
 0 & (2/5,8/15)\\
 1/2 & (4/15,16/25)\ \cup\ (4/5,1]\\
 1 & (1/5,1].
 \end{array}
\]
In $(2/5,8/15)$ the diagnostic gain alone now covers cost. At $V=1/2$
the agents above $4/5$ test solely for the bounty. At $V=1$ agents in
the intervening gap also test. These rules allow us to compare both
state-specific errors as the reward varies.
\end{example}

\section{Learning with and without verification}\label{sec:dynamics}

Return to Example~\ref{ex:disconnected-region}. Take an even prior and let the
signal be correct with probability $q=3/4$. Table~\ref{tab:leading-errors}
distinguishes three outcomes for learning. Agents may obtain proof, which is
possible only in state one. They may also settle on action zero (a lower cascade)
or on action one (an upper cascade), without any further investigation. In state
one, the last of these outcomes gives every subsequent agent the right action even
though some uncertainty persists. We write $\Pr_\theta$ and $\mathbb E_\theta$ for
probabilities and expectations conditional on state $\theta$.

\begin{table}[htbp]
\centering\small
\caption{Terminal outcomes with and without verification}
\label{tab:leading-errors}
\begin{tabular}{@{}lccccc@{}}\toprule
 &\multicolumn{3}{c}{State 1}&\multicolumn{2}{c}{State 0}\\
\cmidrule(lr){2-4}\cmidrule(l){5-6}
 &Proof&Lower (wrong)&Upper&Lower&Upper (wrong)\\\midrule
Actions alone &0&$1/13$&$12/13$&$9/13$&$4/13$\\
Verification available &$0.8585$&$0.1415$&0&1&0\\
\bottomrule
\end{tabular}
\par\smallskip\begin{minipage}{.94\linewidth}\footnotesize
The prior and test success probability are $1/2$, signal precision is
$3/4$, investigation costs $7/10$, and proof pays two. Each state's
columns sum to one. The two decimals are rounded from the bounds
derived in \ref{app:submitted-exits}.
\end{minipage}
\end{table}
\FloatBarrier

Start with state zero, where every investigation fails. Without verification the
first favorable signal can start a wrong upper cascade. With verification a
favorable type is willing to test at every belief in the upper cascade band. An
upper cascade can therefore never begin. In state one the very same decision to
investigate lets some agents obtain proof and lets others see a failure where
previously they would have seen an optimistic action. To account for the second row
of the table, we must follow what the public learns after such a history and
determine where it can stop.

\subsection{Belief transitions}\label{subsec:belief-transitions}

It is convenient to work with the odds on state one. For $y\in(0,1)$, define
\begin{equation}\label{eq:public-odds}
 O(y)\equiv\frac{y}{1-y},
 \qquad
 \Lambda\equiv\frac{q}{1-q}>1.
\end{equation}
Bayes' rule for the private signal and for a failed test can then be written as
\begin{equation}\label{eq:odds-elementary-updates}
\begin{aligned}
 O\!\left(x^1(\mu)\right)&=\Lambda O(\mu),
 &O\!\left(x^0(\mu)\right)&=\Lambda^{-1}O(\mu),\\
 O\!\left(\Phi_p(y)\right)&=(1-p)O(y).
\end{aligned}
\end{equation}
For a perfect test the posterior after any feasible failure is $\Phi_1(y)=0$. We
use $O(0)=0$ in this case.

At a given public belief $\mu$, the only private information that affects the
choice of a test is the signal. Since an indifferent agent abstains, we can write
the decision of type $s$ as
\begin{equation}\label{eq:type-investigation-rule}
 I_s(\mu)
 \equiv
 \mathbf 1\!\left\{x^s(\mu)\in\mathcal X_I(p,c,V)\right\},
 \qquad s\in\{0,1\}.
\end{equation}
Her actions if she abstains and if she tests and fails are, respectively,
\begin{equation}\label{eq:type-action-rules}
 A_s^N(\mu)\equiv a^\star\!\left(x^s(\mu)\right),
 \qquad
 A_s^F(\mu)\equiv
 a^\star\!\left(\Phi_p(x^s(\mu))\right).
\end{equation}
Observers who see her abstain therefore see the record
\[
 r_s^N(\mu)=(0,\varnothing,A_s^N(\mu)),
\]
and observers who see her test and fail see
\[
 r_s^F(\mu)=(1,0,A_s^F(\mu)).
\]
At belief $\mu$ we retain only the record consistent with the type's choice
$I_s(\mu)$ on the branch without success. We say that this record \emph{reveals}
the signal if only one type could have generated it and that it \emph{pools} the
signals if both could have done so. In particular, the public observes the action
after a failure and may use it to distinguish two agents who both chose to test.

\begin{proposition}
\label{prop:exact-public-transition}
At every nonterminal public belief $\mu\in(0,1)$, the next public belief is
\begin{equation}\label{eq:exact-public-transition}
 \mu'=
 \begin{cases}
 1,
 &\text{after a successful investigation},\\[0.35em]
 x^s(\mu),
 &\text{after no investigation when the record reveals $s$},\\[0.35em]
 \mu,
 &\text{after no investigation when the record pools signals},\\[0.35em]
 \Phi_p\!\left(x^s(\mu)\right),
 &\text{after failure when the record reveals $s$},\\[0.35em]
 \Phi_p(\mu),
 &\text{after failure when the record pools signals}.
 \end{cases}
\end{equation}
Consequently, the public belief process is Markov in $\mu$ up to conclusive evidence.
\end{proposition}

\begin{proof}
After success only state~1 is possible. For each of the other records, first ask
which signals can generate it. Their likelihood ratio is $\Lambda^{2s-1}$ if the
record identifies $s$ and one if either signal can generate it. A failed test adds
the factor $1-p$. The outcome factor is one for an abstention. Multiplying the
initial odds by these factors and applying \eqref{eq:odds-elementary-updates} gives
\eqref{eq:exact-public-transition}. Two histories that give the same $\mu$
therefore give the same type decisions and the same distribution of subsequent
beliefs.
\end{proof}

In state~1 a high signal has probability $q$ and a low signal has probability
$1-q$. The chance that the next agent produces proof is consequently
$p[qI_1(\mu)+(1-q)I_0(\mu)]$. It is positive as soon as either type tests. In
state~0 it is zero at every belief.

Fix the public belief and suppose that an agent receives signal $s$. How much would
her action have told the public if a test had been unavailable? In that case the
public would see $A_s^N(\mu)$ alone. Denote the resulting posterior by $b_s(\mu)$.
The distinction matters because in the model with verification an observer who sees
the same action can also see that the agent chose to abstain.

\begin{lemma}\label{lem:selection-versus-failure}
At an interior belief $\mu$, the same-signal action benchmark is
\begin{equation}\label{eq:same-signal-benchmark}
 b_s(\mu)=
 \begin{cases}
 x^s(\mu),&1-q\leq\mu<q,\\
 \mu,&0<\mu<1-q\ \text{or}\ q\leq\mu<1.
 \end{cases}
\end{equation}
The decision to investigate reveals a signal that this action would
conceal if and only if $I_0(\mu)\ne I_1(\mu)$ and $\mu$ lies in one of
the two pooling bands in the second line.

Conditional on an observed attempt, failure multiplies public odds by
$1-p$. If both types test, the attempt leaves beliefs unchanged. For
$p<1$, their actions after failure then reveal the signal if and only if
\begin{equation}\label{eq:post-failure-separation}
 \frac{1}{(1-p)\Lambda}\ \leq\ O(\mu)\
 <\ \frac{\Lambda}{1-p}.
\end{equation}
If only one type tests, the attempt has already identified her signal.

For a complete failed record that identifies $s$, let $\mu'$ be the
posterior after the action. With $\lambda_s=\Lambda^{2s-1}$,
\begin{equation}\label{eq:revealed-signal-failure-odds}
 \frac{O(\mu')}{O(\mu)}=(1-p)\lambda_s,\qquad
 \frac{O(\mu')}{O(b_s(\mu))}=
 \begin{cases}
 1-p,&1-q\leq\mu<q,\\
 (1-p)\lambda_s,&\text{in either pooling band}.
 \end{cases}
\end{equation}
A failed record that pools the signals gives $O(\mu')=(1-p)O(\mu)$.
When $p=1$, a feasible failure reveals state~0 immediately.
\end{lemma}

\begin{proof}
With actions alone, type $s$ chooses one when $\lambda_s O(\mu)\geq1$. At $\mu=1-q$
this includes the high type alone. At $\mu=q$ it already includes both types. This
gives the two boundaries in \eqref{eq:same-signal-benchmark}. When a test is
available, seeing the decision suffices to distinguish the types exactly when their
$I_s(\mu)$ differ. In the lower pooling band this is possible only for
$x^0(\mu)\leq c/\kappa<x^1(\mu)$, with the high type investigating.

Now condition on an attempt. Whichever type made it, its probability of failure is
$1-p$ in state one and one in state zero. Thus failure multiplies the odds already
reached after the attempt by $1-p$. Suppose that both types test and that $p<1$.
The subsequent action identifies the signal when the low type's private odds after
failure are strictly below one and the high type's are at least one. Substitution
from \eqref{eq:odds-elementary-updates} yields \eqref{eq:post-failure-separation}.
Whenever the complete failed record reveals $s$, the combined likelihood ratio is
$(1-p)\lambda_s$. Dividing by the likelihood ratio of the action alone gives
\eqref{eq:revealed-signal-failure-odds}. If the record pools the two types, the
factor due to the signal is one. At $p=1$ the failure has already revealed the
state before the action is observed.
\end{proof}

Suppose an observer sees a failed test and can infer the agent's signal. Relative
to the start of the period, the low signal and the failure both lower the odds. For
the high signal they work in opposite directions. The observer ends the period more
optimistic than she began it precisely when
\begin{equation}\label{eq:positive-selection-condition}
 (1-p)\Lambda>1
 \quad\Longleftrightarrow\quad p<\frac{2q-1}{q}.
\end{equation}
At equality the two pieces of news cancel, and for a higher $p$ the failure
outweighs the signal. Now compare the failed record with the same high type's
action in the absence of a test. If her action alone would have revealed the
signal, the observer would have received the favorable news in any case. She is
therefore less optimistic after the failed test than after that action. If the
failed record pools the types while the action alone would have revealed $s$, the
ratio of the two posterior odds is $(1-p)/\lambda_s$. In the lower action-cascade
band both types take action zero after a failure. There the final action cannot
reveal the signal.

The distinction can be seen in Example~\ref{ex:disconnected-region} with $q=3/4$.
In each row below a high type investigates and fails. The column for her action
alone starts from the same initial public belief as the three columns for the
observed investigation.
\begin{center}
{\small
\setlength{\tabcolsep}{4.5pt}
\begin{tabular}{@{}cccccc@{}}\toprule
 & & \multicolumn{1}{c}{No test available} & \multicolumn{3}{c}{Test available}\\
\cmidrule(lr){3-3}\cmidrule(l){4-6}
Start & Who tests & Action alone & Attempt & Failure & Final action\\
\midrule
$1/4$ & High & $1/2$ & $1/2$ & $1/3$ & $1/3$\\
$1/2$ & High & $3/4$ & $3/4$ & $3/5$ & $3/5$\\
$3/4$ & Both & $3/4$ & $3/4$ & $3/5$ & $9/11$\\
$5/6$ & High & $5/6$ & $15/16$ & $15/17$ & $15/17$\\
\bottomrule
\end{tabular}}
\end{center}
At $\mu=1/2$, the high type's action alone would raise belief to $3/4$. Seeing her
start a test tells the observer exactly as much. Seeing it fail lowers belief to
$3/5$. At $\mu=q$ both types test, so it is their action after failure that
distinguishes them. Action one takes belief from $3/5$ to $9/11$, and action zero
takes it to $1/3$. Finally, at $\mu=5/6$ the two types would both take action one
without a test. Here only the high type investigates. Observing her willingness to
pay reveals a signal that her action alone would have concealed.

\subsection{Why attempts stop}\label{subsec:finite-window}

There is a belief sufficiently low that even an agent with a favorable signal sees
no reason to investigate. To identify such a belief, take
\begin{equation}\label{eq:safe-low-posterior}
 \varepsilon_0=\min\{1/4,c/(2\kappa)\},\qquad
 \delta_0=\psi_1(\varepsilon_0),
\end{equation}
where $\psi_s$ is the inverse of $x^s$. At $\mu\leq\delta_0$ even the high type's
private belief is at most $\varepsilon_0$. Hence $G(x)=\kappa x-c\leq-c/2$ for
either type. Each abstains and takes action zero, leaving the next agent with
exactly the same public belief.

\begin{theorem}\label{thm:finite-investigation}
For $c>0$ and $p>0$, the number of investigations $N_I$ is finite
almost surely in either state. In state one,
\begin{equation}\label{eq:attempt-tail-state-one}
 \Pr_1(N_I\geq n)\leq(1-p)^{n-1}\quad(n\geq1),\qquad
 \mathbb E_1[N_I]\leq1/p .
\end{equation}
The bound at $n=1$ is one, including when $p=1$.
\end{theorem}
\begin{proof}
Consider state one first. Reaching an $n$th investigation requires the first $n-1$
investigations to fail. This gives the stated geometric bound, independently of the
cost. Let $E$ be the event that the public sees infinitely many investigations. It
has probability zero in state one. By posterior convergence, the limiting belief is
$\Pr(\theta=1\mid H_\infty)$ and equals zero almost surely on $E$ (see
\ref{app:martingale-step}). Thus on $E$ the public would eventually reach a belief
below $\delta_0$. At that belief neither type would ever investigate again. It
follows that $E$ also has probability zero in state zero.
\end{proof}

\begin{corollary}\label{cor:pooled-failure-bound}
Start at interior public odds $O_0$. Suppose the next $n$ records are
consecutive failed investigations, each pooling both signals even after
the final action. More generally, allow other nonterminal records
between the starting date and the $n$th such failure if each has
conditional likelihood ratio one. Immediately after the $n$th pooled failure,
\begin{equation}\label{eq:pooled-failure-path}
 O_n=(1-p)^nO_0.
\end{equation}
For $p<1$, no further investigation is possible once
\begin{equation}\label{eq:pooled-failure-bound}
 n\geq
 \left\lceil
 \frac{\log\!\bigl(O_0/O(\delta_0)\bigr)}{-\log(1-p)}
 \right\rceil_+,
 \qquad
 \lceil z\rceil_+\equiv\max\{0,\lceil z\rceil\}.
\end{equation}
If $p=1$, the first failure sets public belief to zero.
\end{corollary}

\begin{proof}
Each time both types produce the same failed record, the odds are multiplied by
$1-p$. Let $L_j$ be the conditional likelihood ratio of any other record between
the starting date and the last such failure. The final odds are
$O_0(1-p)^n\prod_j L_j$. The hypothesis makes each $L_j$ equal to one, which gives
\eqref{eq:pooled-failure-path}. When $p<1$, solving $O_n\leq O(\delta_0)$ for $n$
gives \eqref{eq:pooled-failure-bound}. The next agent then abstains and takes zero
after either signal. When $p=1$ the first failure establishes state~0.
\end{proof}

\subsection{False-cascade traps}\label{subsec:false-traps}

If agents who receive different signals take the same action, their predecessors'
actions may appear to have started a cascade. Yet observers might still distinguish
the two types by which one investigates. We therefore ask when the \emph{entire}
public record repeats.

\begin{definition}\label{def:cascade-trap}
A nonterminal public belief $\mu$ is a cascade trap for action $a\in\{0,1\}$ if both signal types decline investigation and choose action $a$. Let $\mathcal T_0$ and $\mathcal T_1$ denote the sets of lower and upper traps. If $\theta\neq a$, the corresponding trap is a false cascade.
\end{definition}

At such a belief, both types generate $(0,\varnothing,a)$. This record has
probability one in either state and leaves the belief at $\mu$. The same reasoning
applies at every subsequent date.

\begin{proposition}\label{prop:exact-trap-sets}
The action tie-breaking rule $a^\star(1/2)=1$ implies
\begin{align}
 \mathcal T_0
 &=\left\{\mu\in(0,1-q):
 x^1(\mu)\leq\frac{c}{\kappa}\right\},
 \label{eq:lower-trap-set}\\
 \mathcal T_1
 &=\left\{\mu\in[q,1):
 G(x^0(\mu))\leq0
 \ \text{and}\
 G(x^1(\mu))\leq0\right\}.
 \label{eq:upper-trap-set}
\end{align}
In particular,
\begin{equation}\label{eq:lower-trap-explicit}
 \mathcal T_0=
 \begin{cases}
 \bigl(0,\psi_1(c/\kappa)\bigr],
 &c<c_M=\kappa/2,\\[0.35em]
 (0,1-q),
 &c\geq c_M.
 \end{cases}
\end{equation}
Thus a nonempty lower trap exists for every $c>0$.
\end{proposition}

\begin{proof}
In the lower band the high type has the larger gain from testing, since
$G(x)=\kappa x-c$ for both types. As soon as she is willing to abstain, so is the
low type. In the upper band abstention gives action one for both types, and each
must have a nonpositive testing gain. For beliefs between the bands, different
signals induce different actions if both types abstain.
\end{proof}

\begin{corollary}\label{cor:upper-trap-intervals}
If the private testing set is $(\ell,u)\cup(h,1]$ with
$\ell<1/2<u\leq h<1$, then
\begin{equation}\label{eq:upper-trap-gap}
 \mathcal T_1=\{\mu:\ \psi_0(u)\leq\mu\leq\psi_1(h)\}.
\end{equation}
It is empty when the two bounds cross. For an upper testing interval
$(r,1]$ an upper trap instead requires $x^1(\mu)\leq r$ at $\mu\geq q$.
For a bounded central interval $(\ell,u)$ it requires
$x^0(\mu)\geq u$ at $\mu\geq q$. If no type ever tests,
$\mathcal T_1=[q,1)$.
\end{corollary}
\begin{proof}
If $\mu\geq q$, then $1/2\leq x^0(\mu)<x^1(\mu)$. With a split testing set,
abstention by both types requires $x^0(\mu)\geq u$ and $x^1(\mu)\leq h$. Inverting
gives \eqref{eq:upper-trap-gap}. For a single testing interval only the relevant
one of the two boundary tests is needed.
\end{proof}

\begin{corollary}\label{cor:failure-triggered-trap}
After a failed test the public enters a permanent cascade exactly when
the posterior from the complete record belongs to
$\mathcal T_0\cup\mathcal T_1$, for $p<1$. In odds the target is
$(1-p)O(\mu)$ for a pooled record and $(1-p)\Lambda^{2s-1}O(\mu)$
if the record identifies $s$. At $p=1$ failure establishes state~0.
\end{corollary}
\begin{proof}
Proposition~\ref{prop:exact-public-transition} gives the odds after the record.
Applying Definition~\ref{def:cascade-trap} at these odds gives the claim. See
\ref{app:trap-boundaries} for the inequalities at the endpoints.
\end{proof}

\subsection{Terminal outcomes}\label{subsec:terminal-outcomes}

Knowing that the number of investigations is finite leaves open a question about
actions. After the last test there can still be dates at which the action reveals
the signal. We now show that the public reaches proof or an absorbing belief at a
finite date, and bound the expected number of agents who act before then.

Hold the posterior fixed after a conclusive finding. Let $\tau$ be the number of
agents who have acted when the posterior first enters
$\{0,1\}\cup\mathcal T_0\cup\mathcal T_1$. A prior in a trap therefore has
$\tau=0$. We use $L$ and $U$ for the events of first entry into the interior lower
and upper traps $\mathcal T_0$ and $\mathcal T_1$.

\begin{theorem}\label{thm:terminal-outcomes}
Fix a common cost $c>0$, a reward $V\geq0$, signal precision
$1/2<q<1$, and an interior prior $\pi$.
\begin{enumerate}[label=(\roman*),leftmargin=2.2em]
\item If $0<p<1$, exactly one of the following occurs at a finite date
almost surely under either state. A success proves state~1, or the process
enters a lower or an upper trap. In state~0 the two
trap probabilities sum to one.
\item If $p=1$ and a test occurs, it reveals the state, through success
in state~1 and failure in state~0. On histories without a test an
interior trap is reached at a finite date almost surely.
\item For every $0<p\leq1$, there is a finite integer $m\geq1$,
depending on the primitives and the prior, such that, for each
$j=0,1,\ldots$,
\begin{equation}\label{eq:calendar-tail}
 \Pr_\theta(\tau>jm)\leq (1-w_\theta^{m})^j,
 \qquad \mathbb E_\theta[\tau]\leq m/w_\theta^m,
 \qquad w_0=q,\quad w_1=1-q .
\end{equation}
\end{enumerate}
\end{theorem}

\begin{proof}
We first show that the odds before absorption have a common upper bound.
Sufficiently close to one, the private beliefs of both types lie on the same affine
piece of $G$, on which the sign of the gain is constant in a left neighborhood of
one. Thus both types make the same testing decision there. Choose $M$ large enough
that above odds $M$ this is true, both untested actions are one, and, for $p<1$,
both actions after failure are also one. If the types abstain there, the belief is
an upper trap. If they test, success ends the process and failure multiplies odds
by $1-p$, with conclusive failure at $p=1$. From odds at or below $M$, an
inconclusive record can raise odds by at most a factor $\Lambda$. Hence
$K=\max\{O(\pi),\Lambda M\}$ bounds every reachable interior odds value.

Now give the next agent a low signal at any belief outside a trap. If she abstains,
her record must reveal that signal, since pooled abstention with the same action
would itself be a trap. The odds are divided by $\Lambda$. If she tests and the
outcome is inconclusive, the multiplier is at most $1-p$, even when the failure
pools. Each low signal before absorption therefore ends learning or multiplies the
odds by at most $b=\max\{\Lambda^{-1},1-p\}<1$. Take $m\geq1$ such that
$b^mK\leq O(\delta_0)$. Wherever a block of $m$ low signals starts, it forces
absorption by its end. The probability of such a block under state $\theta$ is
$w_\theta^m$ independently of the past. Applying this bound to successive blocks
and summing the tail probabilities gives \eqref{eq:calendar-tail}. For $p<1$, any
feasible finite record without proof is possible in both states, so the only other
outcomes are the two interior traps. For $p=1$ the first test, if there is one,
reveals the state.
\end{proof}

With an imperfect test, even a large bounty can leave state one unproved. We can
bound the probability of a wrong lower cascade in that state from the probability
of lower cascades in state zero.

\begin{corollary}\label{cor:imperfect-error}
Suppose $0<p<1$. From every interior prior there is positive probability
in state~1 that conclusive evidence is never produced. For the two
state-conditional error probabilities,
\begin{equation}\label{eq:floor-on-errors}
 \Pr_1(L)\ \geq\ \rho_\pi\,[1-\Pr_0(U)],\qquad
 \rho_\pi=\min\!\left\{1,\,
       \frac{(1-p)O(\delta_0)}{\Lambda O(\pi)}\right\}>0.
\end{equation}
If $\pi<q$, including $\pi=1/2$, a wrong lower cascade has strictly
positive probability in state~1 for every $c>0$ and every finite $V\geq0$.
\end{corollary}

\begin{proof}
Fix a finite public history $h$ that first reaches a trap and write $O_h$ for the
odds at entry. Its probabilities in the two states satisfy
$\Pr_1(h)=\Pr_0(h)O_h/O(\pi)$. When the prior is a trap, $O_h=O(\pi)$. Otherwise
the odds immediately before entry exceed $O(\delta_0)$, and even the smallest
multiplier from an inconclusive record is at least $(1-p)/\Lambda$. In either case
$O_h/O(\pi)\geq\rho_\pi$. These first-entry histories are disjoint and have total
mass one in state~0. Adding their state-one probabilities gives
$\Pr_1(L\cup U)\geq\rho_\pi$. Restricting the sum to lower entries and using
$\Pr_0(L)=1-\Pr_0(U)$ yields \eqref{eq:floor-on-errors}.

For the last claim, use the block of length $m$ from
Theorem~\ref{thm:terminal-outcomes}. Draw in advance the signals and potential test
outcomes of these $m$ agents. The event of $m$ low signals and $m$ potential
failures has probability $[(1-q)(1-p)]^m>0$ in state~1. No investigation on this
path can produce proof, and the odds decrease whenever the process continues. If
$\pi<q$, the whole path stays below the upper band $[q,1)$ and must end in $L$.
\end{proof}

In particular, if $0<p<1$ and the public cannot reach an upper trap, it enters the
lower trap almost surely in state zero. Equation~\eqref{eq:floor-on-errors} then
puts a strictly positive lower bound on the probability of that same cascade in
state one. An upper trap consisting of a single belief affects this calculation
only if some history hits that belief. To see why the restriction to imperfect
tests matters, set $p=1$, $V=0$, and $\pi=1/2$. Either signal type would gain $1-q$
in accuracy from testing. For $0<c<1-q$ the first agent investigates and
immediately establishes the true state.

\subsection{How verification changes the probability of a wrong cascade}
\label{subsec:absorption-probabilities}

To return to Table~\ref{tab:leading-errors}, we first derive a recursion for the
probability of each terminal outcome in each state. We then use the actions-alone
benchmark to compare the two errors at the cost and bounty of
Example~\ref{ex:disconnected-region} for every signal precision. Finally, we ask
how the state-one comparison changes when the bounty varies.

\smallskip\noindent\emph{Computing the probabilities.}
Starting from a public belief $\mu$, denote by $h_{\theta,A}(\mu)$ the probability
of terminal outcome $A$ conditional on state $\theta$. The outcomes are a lower
cascade $L$, an upper cascade $U$, or proof of state~1, $D_1$. A perfect test can
also produce proof of state~0, $D_0$. At its own outcome each function takes the
value one and at the other outcomes it takes the value zero. Consider a type $s$ at
a transient belief, and write $f_s$ for the posterior that her complete record
produces if there is no success. Proposition~\ref{prop:exact-public-transition}
tells us when to condition this posterior on her signal. At $p=1$ set $f_s=0$ if
she tests. Dropping the argument $\mu$ from $I_s$ and $f_s$, let $b_{\theta s}=q$
if $s=\theta$ and $1-q$ otherwise. Conditioning on the signal and the outcome gives
\begin{align}
 h_{0,A}(\mu)&=\sum_{s=0}^1 b_{0s}\,h_{0,A}(f_s),
 \label{eq:absorption-state-zero}\\
 h_{1,A}(\mu)&=\sum_{s=0}^1 b_{1s}
 \left[(1-pI_s)h_{1,A}(f_s)+pI_s\mathbf 1\{A=D_1\}\right].
 \label{eq:absorption-state-one}
\end{align}
Iterate the equations through $n$ agents. The boundary values determine the
contribution of every history absorbed by then. The difference between two bounded
solutions at the initial belief can therefore be at most their supremum difference
times $\Pr_\theta(\tau>n)$. This last probability tends to zero by
Theorem~\ref{thm:terminal-outcomes}, so the bounded solution is unique. The same
argument gives a way of reading a finite calculation. The mass already at an
outcome is a lower bound on its probability, and adding all the mass still active
gives an upper bound. Since the first success ends investigation, summing the
success probabilities over attempts also yields $\Pr_1(D_1)=p\,\mathbb E_1[N_I]$.

\smallskip\noindent\emph{The benchmark without verification.}
Let $u_\theta(\pi)$ be the probability of an upper cascade when verification is
unavailable. With $D=1-q(1-q)$ and $b=\Pr_\theta(s=1)$, this probability is
\begin{equation}\label{eq:benchmark-all-priors}
 u_\theta(\pi)=
 \begin{cases}
 0,&\pi<1-q,\\
 b^2/D,&1-q\leq\pi<1/2,\\
 b/D,&1/2\leq\pi<q,\\
 1,&q\leq\pi .
 \end{cases}
\end{equation}
An action alone reveals the signal at odds in $[\Lambda^{-1},\Lambda)$. Starting
from any odds in this band, exactly two points of the resulting grid lie inside it.
Call the upper-cascade probabilities at the lower and higher points $u_-$ and
$u_+$. From the lower point the public needs a high signal simply to get back to
the higher point, so $u_-=b u_+$. From the higher point one high suffices to start
the upper cascade, so $u_+=b+(1-b)u_-$. Solving gives
\eqref{eq:benchmark-all-priors}. The inclusion of the left endpoint follows from
the action tie. In particular, at $\pi=1/2$ the wrong-cascade probabilities are
$(1-q)^2/D$ in state one and $(1-q)/D$ in state zero.

Consider the symmetric prior and suppose that $p<1$ and no history reaches the
upper trap. We then have $\Pr_0(L)=1$. The identity in the proof of
Corollary~\ref{cor:imperfect-error}, summed over first entries, gives
$\Pr_1(L)=\mathbb E_0[O(\mu_{\tau+1})]$. Every lower entry has odds strictly below
$1/\Lambda$. With equal prior weights, the error is therefore strictly below
$1/(2\Lambda)$. The benchmark error is
$(\Lambda+2)/[2(1+\Lambda+\Lambda^2)]>1/(2\Lambda)$. Thus eliminating wrong upper
cascades improves average terminal accuracy even if verification makes a mistake
more likely in the other state.

\begin{proposition}\label{prop:submitted-error-comparison}
In Example~\ref{ex:disconnected-region}, fix $p=1/2$, $V=2$, $c=7/10$
and $\pi=1/2$. For every $q\in(1/2,1)$, access to verification strictly
raises the probability of a wrong lower cascade in state~1 and eliminates
wrong upper cascades in state~0. The ex ante probability of a wrong terminal
action falls.
\end{proposition}

To see the two effects at $q=3/4$, start with the state where no proof is possible.
With actions alone the errors are $4/13$ in state zero and $1/13$ in state one.
Anywhere in the upper cascade band the high type assigns probability at least
$9/10$ to state one. Her expected bounty alone exceeds the cost $7/10$, so she
investigates. There is no upper trap. By Theorem~\ref{thm:terminal-outcomes}, state
zero ends in the lower trap with probability one.

In state one it is enough to follow two histories. Two initial lows take odds from
one to $1/3$ and then to $1/9$. This has probability $1/16$. Alternatively, let the
first agent receive a high signal and fail to produce proof. Odds rise to $3/2$.
Two lows now take them to $1/2$ and then to $1/6$. This history has probability
$3/128$. The two histories are disjoint and both end in the lower trap. They
already account for $11/128>1/13$, which proves the increase in error. Iterating
\eqref{eq:absorption-state-one} gives the value in Table~\ref{tab:leading-errors}.
All other eventual state-one mass goes to proof. With equal prior weights, the
error falls from $5/26$ to about $0.0707$. The proof in \ref{app:submitted-exits}
treats every $q$, including those for which an upper trap exists but cannot be
reached from the prior.

\smallskip\noindent\emph{Changing only the bounty.}
We can say more about the role of the bounty if a high signal and a failure exactly
offset one another in odds, that is, if $(1-p)\Lambda=1$. Starting at even odds,
write $\Lambda^k$ for public odds and $\Lambda^j$ for the agent's private odds. If
a failed record reveals a high signal, the next agent starts at precisely the
public odds faced by her predecessor. This lets us identify the entry belief for
each reachable cascade and obtain both of its state-conditional probabilities.

\begin{theorem}\label{thm:bounty-exits}
Let $\pi=1/2$, $1-p=\Lambda^{-1}$, and
$t_j=(c/p)(1+\Lambda^{-j})$ for $j\geq1$.
If $c<p(1+V)/2$, put $C=p(1+V)/c-1$ and define
\[
 k_L=-1-\lceil\log_\Lambda C\rceil,\qquad
 u=(1-q)^2\,\mathbf 1\{V\leq t_3\}.
\]
The absorption probabilities are
\begin{align}
 \Pr_0(U)&=u, &\Pr_0(L)&=1-u,\label{eq:closed-state-zero}\\
 \Pr_1(U)&=\Lambda^2u, &\Pr_1(L)&=\Lambda^{k_L}(1-u).
 \label{eq:closed-state-one}
\end{align}
The remaining state-1 mass is proof.

If $c\geq p(1+V)/2$, the outcomes coincide with the no-verification
benchmark when $V\leq t_2$. When $V>t_2$, state~0 reaches the lower trap
with probability one. In state~1 the lower trap has probability
$\Lambda^{-2}$ and the remaining probability goes to proof.
\end{theorem}

For the first case, consider the last record before entry. Lower entry can occur
only at odds $\Lambda^{k_L}$. If the upper trap is reachable, the first two signals
must be high and the entry odds are $\Lambda^2$. In state zero the corresponding
probabilities are $1-u$ and $u$. Since the prior odds are one, multiplying the
state-zero mass of each first-entry history by its terminal odds gives its mass in
state one. This gives \eqref{eq:closed-state-one}. In the second case the agent
with private belief one half abstains, and the only possible lower entry is at
$\Lambda^{-2}$. The derivation in \ref{app:submitted-exits} includes the equality
cases at the testing cutoffs.

Take $q=2/3$, $p=1/2$, and $c=1/5$. Even at a zero bounty the midpoint type prefers
to test, so the first case of Theorem~\ref{thm:bounty-exits} applies for all
$V\geq0$. Table~\ref{tab:terminal-comparison} gives the outcomes over the first
four bounty intervals. For comparison it also gives the outcome with the cost and
bounty of Example~\ref{ex:disconnected-region} at the same $q$.
\begin{table}[htbp]
\centering\small
\caption{Terminal probabilities at $q=2/3$, $p=1/2$ and $\pi=1/2$.}
\label{tab:terminal-comparison}
\begin{tabular}{@{}lccccc@{}}\toprule
 & \multicolumn{3}{c}{State 1} & \multicolumn{2}{c}{State 0}\\
\cmidrule(lr){2-4}\cmidrule(l){5-6}
Regime & Proof & Lower & Upper & Lower & Upper\\\midrule
Verification unavailable & 0 & $1/7$ & $6/7$ & $4/7$ & $3/7$\\
$c=7/10$, $V=2$ & $3/4$ & $1/4$ & 0 & 1 & 0\\
\addlinespace
\multicolumn{6}{@{}l}{Common cost $c=1/5$}\\
$0\leq V\leq 1/5$ & $1/3$ & $2/9$ & $4/9$ & $8/9$ & $1/9$\\
$1/5<V\leq 9/20$ & $4/9$ & $1/9$ & $4/9$ & $8/9$ & $1/9$\\
$9/20<V\leq 1$ & $7/8$ & $1/8$ & 0 & 1 & 0\\
$1<V\leq 13/5$ & $15/16$ & $1/16$ & 0 & 1 & 0\\\bottomrule
\end{tabular}
\end{table}

At this cost, verification raises the state-one error above the benchmark $1/7$ for
$0\leq V\leq1/5$ and lowers it for every $V>1/5$. Within the latter range, consider
increasing $V$ just beyond $9/20$. The change concerns a favorable type whose
predecessors have already taken public odds to four. At the lower bounty she would
have abstained, leaving the public in an upper cascade. In state~1 that cascade was
correct and had probability $4/9$. At the higher bounty she tests. Starting from
odds four, the probability of reaching the wrong lower cascade is now
$(1/8)/4=1/32$. Consequently the increase in the initial state-one error is
$(4/9)(1/32)=1/72$. The rest of the mass previously assigned to the correct cascade
goes to proof.

Figure~\ref{fig:bounty-errors} plots the first three jumps. Above $9/20$ further
changes occur only when agents start testing at a lower belief and the lower entry
moves down by another power of two. Its state-one probability tends to zero as the
bounty grows. The ordering between three interior bounties and the no-test
benchmark persists on an open set where the two precisions vary independently
(\ref{app:off-grid}).

\begin{figure}[htbp]
\centering
\begin{tikzpicture}[
  x=7.5cm,y=25.5cm,
  font=\small,
  axis/.style={black!75,line width=0.45pt},
  benchmark/.style={black!70,dash pattern=on 3pt off 2pt,line width=0.65pt},
  probability/.style={black,line width=1.15pt,line cap=butt},
  ticklabel/.style={font=\scriptsize,inner sep=1pt}
]
  \def\vmax{4/3}
  \draw[axis] (0,0) -- ({\vmax},0);
  \draw[axis] (0,0) -- (0,.24);
  \node[anchor=south west,inner sep=0pt] at (0,.247)
    {$\Pr_1(L)$};

  \foreach \x/\lab in {0/0,{1/5}/{1/5},{9/20}/{9/20},1/1,{4/3}/{4/3}}{
    \draw[axis] ({\x},0) -- ++(0,-2.0pt);
    \node[ticklabel,anchor=north,yshift=-3pt] at ({\x},0) {$\lab$};
  }
  \node[anchor=north east,inner sep=0pt,yshift=-14pt] at ({\vmax},0)
    {Bounty $V$};
  \foreach \y/\lab in {{1/16}/{1/16},{1/9}/{1/9},{1/8}/{1/8},{2/9}/{2/9}}{
    \draw[axis] (0,{\y}) -- ++(-2.0pt,0);
    \node[ticklabel,anchor=east,xshift=-3pt] at (0,{\y}) {$\lab$};
  }

  \draw[benchmark] (0,{1/7}) -- ({\vmax},{1/7});
  \node[anchor=south east,inner sep=1pt,yshift=1pt,font=\scriptsize]
    at ({\vmax},{1/7}) {No verification $\;(1/7)$};

  \draw[probability] (0,{2/9}) -- ({1/5},{2/9});
  \draw[probability] ({1/5},{1/9}) -- ({9/20},{1/9});
  \draw[probability] ({9/20},{1/8}) -- (1,{1/8});
  \draw[probability] (1,{1/16}) -- ({\vmax},{1/16});
  \foreach \x/\y in {0/{2/9},{1/5}/{2/9},{9/20}/{1/9},1/{1/8}}
    \fill[black] ({\x},{\y}) circle[radius=2pt];
  \foreach \x/\y in {{1/5}/{1/9},{9/20}/{1/8},1/{1/16}}
    \draw[black,fill=white,line width=.65pt] ({\x},{\y}) circle[radius=2pt];
\end{tikzpicture}
\caption{A wrong lower cascade in state~1 as the bounty varies.
Here $q=2/3$, $p=\pi=1/2$ and $c=1/5$. Filled circles give the value at a cutoff.}
\label{fig:bounty-errors}
\end{figure}
\FloatBarrier

\subsection{What sustains testing at optimistic beliefs?}
\label{subsec:reward-and-gap}

Two features of the testing decision matter at optimistic beliefs. A favorable type
may pay to establish something she already takes to be true. We first let the
payment decline when the finding is widely anticipated. Even with a fixed bounty
the agent may decline at some private beliefs between two regions in which she
tests. If both signals put the agent in this gap, an optimistic cascade can begin.
To see what the gap contributes, we construct two policies that differ only in the
testing decisions made inside it.

\smallskip\noindent\emph{A reward for a surprising finding.}
Let $R(\mu)$ be the amount paid to the agent who produces proof, where $\mu$ is the
public belief immediately before her investigation decision. The sponsor announces
the rule in advance. With a fixed $V$, the investigator earns the same bounty
whether proof comes as a surprise or confirms what everyone expected. If part of
the return is the credit an investigator receives for an unexpected finding, a
natural alternative is $A(1-\mu)$, $A\geq0$. We use this payment as a reduced-form
specification of that return. The net gain from testing for an agent with signal
$s$ is
\begin{equation}\label{eq:public-reward-gain}
 D_p(x^s(\mu))+p\,x^s(\mu)R(\mu)-c .
\end{equation}
In particular, the decision to start the test may itself raise the public belief.
The investigator's promised payment still depends on the belief before that
decision. Observers update on her decision, the test result, and the action as in
Section~\ref{subsec:belief-transitions}.

\begin{proposition}\label{prop:surprise-exits}
Let $c>0$. For any bounded nonnegative reward $R$ with
$\lim_{\mu\uparrow1}R(\mu)=0$, all sufficiently optimistic interior
beliefs are upper traps. The finite-time classification of
Theorem~\ref{thm:terminal-outcomes} continues to hold.

For $R(\mu)=A(1-\mu)$, suppose $\pi=1/2$, $(1-p)\Lambda=1$, and
$c<p/2$. Then
\begin{equation}\label{eq:taper-upper}
 \Pr_0(U)=(1-q)^2\,
 \mathbf 1\!\left\{A\leq\frac{2c}{p}(1+\Lambda^{-1})\right\},
 \qquad \Pr_1(U)=\Lambda^2\Pr_0(U).
\end{equation}
\end{proposition}

When $\mu$ is sufficiently close to one, even the reward for a successful test is
small and neither type gains much from the information. Both decline and take
action one. At the other end, a bounded payment cannot cover the cost when the
probability of success is sufficiently small. The block of low signals used in
Theorem~\ref{thm:terminal-outcomes} therefore still forces absorption. For the
linear schedule, what determines whether the upper trap can be reached from even
odds is the favorable type's decision at those initial odds. If she declines there,
precisely two initial highs are needed for upper entry. If she tests there, the
upper trap is unreachable. The proof in \ref{app:reward-comparisons} derives
\eqref{eq:taper-upper} and the corresponding lower-exit probability.

To compare payments of the same size at the prior, keep $(1-p)\Lambda=1$, $c<p/2$,
and $\pi=1/2$, and compare a fixed $V$ with $R(\mu)=2V(1-\mu)$.
Theorem~\ref{thm:bounty-exits} and Proposition~\ref{prop:surprise-exits} give the
range in which the upper cascade disappears under the fixed payment and remains
reachable under the declining payment. The condition is
\[
 \frac{c}{p}(1+\Lambda^{-3})<V\leq
 \frac{c}{p}(1+\Lambda^{-1}).
\]
Even under the declining payment, a sufficiently high $A$ prevents the public from
getting from the prior to the upper neighborhood in which both types would abstain.

For instance, take $p=\pi=1/2$, $q=2/3$, $c=1/5$, and fixed $V=1/2$ as in
Table~\ref{tab:terminal-comparison}. Under the fixed bounty and the two declining
payments $(1-\mu)/2$ and $1-\mu$, both types make the same decisions until odds
reach four. To reach four for the first time requires two initial high signals.
There the two private posteriors are $2/3$ and $8/9$. Either type would choose
action~1 even after failure. For the fixed payment the expected bounties are $1/6$
and $2/9$, so only the high type investigates. Now consider the more generous of
the two declining schedules. At these odds it pays $1/5$ for proof, worth only
$4/45$ in expectation even to the high type. She too abstains, and an upper cascade
begins.

In state~0 this change raises the probability of the wrong upper cascade from zero
to $1/9$. In state~1 the probabilities of proof, lower entry, and upper entry
change from $(7/8,1/8,0)$ to $(4/9,1/9,4/9)$. The same histories and probabilities
obtain for $1/5<c<5/24$, where every investigation decision is strict. See
\ref{app:reward-comparisons} for the cutoffs and the other exit probabilities for
general $A$.

\smallskip\noindent\emph{What the gap changes.}
Return to a fixed payment at the parameters of Example~\ref{ex:two-motives}.
Raising the bounty to $V=1$ makes the testing region a single interval reaching
certainty. The upper trap is still empty and the state-one probabilities are again
$(7/8,1/8,0)$. Thus testing at high beliefs can eliminate the upper trap under
either a connected or a disconnected rule. For a gap in testing to sustain an upper
cascade it must accommodate both private posteriors at a public belief that can
actually be reached. A narrower gap can still induce one of the two types to
abstain.

Take a testing region $(\ell,u)\cup(h,1]$ with $\ell<1/2<u\leq h<1$. At public odds
$O$, the private odds after low and high signals are $O/\Lambda$ and $\Lambda O$.
To sustain an upper cascade, both corresponding posteriors must lie in $[u,h]$ by
Corollary~\ref{cor:upper-trap-intervals}. Expressed in public odds, the interval is
\begin{equation}\label{eq:gap-in-odds}
 [\Lambda O(u),O(h)/\Lambda],\qquad
 \text{nonempty exactly when}\quad O(h)/O(u)\geq\Lambda^2 .
\end{equation}
At equality there is only one such belief. Entry from a given prior then requires
some history to hit it exactly.

\begin{proposition}\label{prop:gap-filling}
For any split testing set $(\ell,u)\cup(h,1]$ in this model, there is a
fixed bounty and common cost, at the same test quality, that induce
$(\ell,1]$. The only private beliefs at which the agent's testing choice
changes are those in $[u,h]$.

At $p=1/2$, $q=4/7$ and $\pi=2/3$, start from $(V,c)=(1/2,1/5)$.
Filling its gap by changing $(V,c)$ to $(1,4/15)$ removes a wrong upper
cascade of probability $3/7$ in state~0. In state~1 it raises the
probability of proof by $51/98$ and the probability of a wrong lower
cascade by $5/98$.
\end{proposition}

\begin{proof}
Set $V'=2/p-1$ and $c'=2\ell$, so that $p(1+V')=2$. The gain first becomes positive
above $\ell$, stays at $1-2\ell>0$ on the middle piece, and is increasing on the
last one. Thus an agent tests exactly on $(\ell,1]$. Relative to the original rule
the change is confined to $[u,h]$, including the indifferent endpoints.

In the example the original testing set is $(4/15,16/25)\cup(4/5,1]$. We can also
obtain $(4/15,1]$ with $V'=1$ and $c'=4/15$ by
Theorem~\ref{thm:investigation-geometry}. It is the adjustment in cost that keeps
the lower cutoff fixed. Give the first agent the low signal. Her private odds are
$3/2$, so she tests under either policy. Give her the high signal instead. Her
private odds $8/3$ lie in the original gap and she abstains under that rule.
Observers infer her signal and revise public odds from two to $8/3$. As
Figure~\ref{fig:gap-mechanism} shows, even the two possible private posteriors of
the \emph{next} agent now fit in the gap.

Follow the initial low branch first. Under both rules a failed record identifies
the low type and takes public odds to $3/4$. From there even the high type's
private posterior is at most $1/2$. At a transient belief in this range the high
type tests, and any inconclusive record multiplies the odds by at most
$\max\{\Lambda^{-1},(1-p)\Lambda\}=3/4$. Both policies therefore agree throughout
this continuation. Under the original rule, upper entry occurs precisely on an
initial high signal.

Now give the initial agent a high signal under the policy that fills the gap. She
tests and takes action~1 after a failure. The action identifies her signal and
leaves the public at odds $4/3$. In the next two periods, at odds $4/3$ and $2/3$,
both types test and pool after a failure. Two failures therefore take odds to
$1/3$. There a low type abstains and a high type tests. Either a low or a high
followed by failure takes the public into the lower trap. Conditional on an initial
high and state~1, the probability of that exit is consequently
\[
 (1/2)^3\big[(3/7)+(4/7)(1/2)\big]=5/56 .
\]
The conditional probability of proof is the remaining $51/56$. The initial high has
probability $4/7$ in state~1, so weighting the two continuations gives the
increases in error and proof in the statement. In state~0 the same initial high now
leads to a lower cascade.
\end{proof}

One can also vary the bounty at the original cost $c=1/5$. A bounty of one
eliminates the upper trap. At $V=8/15$ the testing set still has two components,
but the private beliefs after the two signals are too far apart to fit in its gap.
Alternatively, keep $V=1/2$. An initial high then starts the upper cascade for all
$8/41\leq c<1/4$, including the lower endpoint where the next favorable type is
indifferent. The derivations are in \ref{app:reward-comparisons}. It is in these
probabilities that the shape of the testing set has consequences for learning. The
updating argument in Lemma~\ref{lem:selection-versus-failure} requires knowing
which type produced the observed record. The stopping argument in
Theorem~\ref{thm:finite-investigation} uses the chance of success and the positive
cost. Each argument applies to either shape of testing set.

\begin{figure}[htbp]
\centering
\begin{tikzpicture}[
  x=2.18cm,y=0.98cm,font=\small,
  rowname/.style={anchor=east,inner sep=0pt},
  track/.style={black!45,line width=.35pt},
  tests/.style={black,line width=1.2pt},
  marklabel/.style={font=\scriptsize,inner sep=1pt,anchor=south},
  bound/.style={draw=black,fill=white,line width=.7pt}
]
  \def\leftend{.16}
  \def\rightend{4.48}
  \def\entry{4/11}
  \def\gapstart{16/9}

  \fill[black!7] ({\gapstart},.22) rectangle (4,3.06);
  \node[font=\scriptsize,inner sep=0pt,anchor=south]
    at ({(16/9+4)/2},3.09) {Gap in the original rule};

  \foreach \y/\name in {4/{Gap filled},2.78/{Original rule}}{
    \node[rowname] at (.07,\y) {\name};
    \draw[track] (\leftend,\y) -- (\rightend,\y);
  }
  \draw[tests,->] ({\entry},4) -- (\rightend,4);
  \draw[tests] ({\entry},2.78) -- ({\gapstart},2.78);
  \draw[tests,->] (4,2.78) -- (\rightend,2.78);
  \foreach \x/\y in {{4/11}/4,{4/11}/2.78,{16/9}/2.78,4/2.78}
    \draw[bound] ({\x},\y) circle[radius=2pt];

  \foreach \y/\name in {1.73/{Prior $O=2$},.67/{Next $O=8/3$}}{
    \node[rowname] at (.07,\y) {\name};
    \draw[track] (\leftend,\y) -- (\rightend,\y);
  }
  \foreach \x/\y/\what/\lab in {
       {3/2}/1.73/{low}/{3/2},
       {8/3}/1.73/{high}/{8/3},
       2/.67/{low}/{2},
       {32/9}/.67/{high}/{32/9}}{
    \node[marklabel,yshift=3pt] at ({\x},\y) {\what\ $\lab$};
  }
  \fill ({3/2},1.73) circle[radius=2.2pt];
  \foreach \x/\y in {{8/3}/1.73,2/.67,{32/9}/.67}
    \draw[bound] ({\x},\y) circle[radius=2.2pt];

  \draw[->,black!75,line width=.45pt] (\leftend,.04) -- (\rightend,.04);
  \foreach \x/\lab in {{4/11}/{4/11},{16/9}/{16/9},4/4}{
    \draw[black!75,line width=.45pt] ({\x},.04) -- ++(0,-2.1pt);
    \node[anchor=north,font=\scriptsize,inner sep=1pt,yshift=-2pt]
      at ({\x},.04) {$\lab$};
  }
  \node[anchor=north,font=\small,inner sep=0pt] at (2.25,-.38) {Private odds};
  \node[anchor=west,font=\scriptsize,inner sep=0pt] at (.30,3.52)
    {Heavy segments indicate testing.};
\end{tikzpicture}

\caption{The two private signals before and after a revealing abstention.
In the signal rows, filled markers denote testing and open markers denote
abstention under the original rule. The connected rule fills the shaded gap.}
\label{fig:gap-mechanism}
\end{figure}
\FloatBarrier

\section{Robustness and extensions}\label{sec:robustness}

So far, an agent who sees a positive test knows the state, and an observer who
knows the signal knows exactly whether the agent will test. We relax these
assumptions in turn. First a positive finding can be mistaken, so the next agent
may still wish to investigate. Then we return to conclusive proof and let two
agents with the same signal face different costs. In that case even an abstention
followed by a common action may be informative.

\subsection{Noisy verification and false positives}
\label{subsec:noisy-verification}

Suppose that an investigation produces a public finding $z\in\{+,-\}$ with
probabilities
\begin{equation}\label{eq:noisy-verification-technology}
 \Pr(z=+\mid\theta=1)=p,
 \qquad
 \Pr(z=+\mid\theta=0)=\eta,
 \qquad
 0\leq\eta<p\leq1.
\end{equation}
The sponsor pays $V$ whenever the finding is positive, including when it is false.
If $0<\eta<p<1$, agents continue to act after either finding and a positive today
does not exhaust the bounty for tomorrow. At $\eta=0$ a positive proves state~1 and
ends the baseline game. At $p=1$ a negative proves state~0. We first derive the
private decision, then study continued investigation when either finding is
possible in both states.

\begin{proposition}\label{prop:noisy-gain}
Write $D_{p,\eta}(x)$ for the expected increase in accuracy from
conditioning the action on the test outcome. The net gain is
\begin{equation}\label{eq:noisy-gain-decomposition}
 G_\eta(x)=D_{p,\eta}(x)+V[\eta+(p-\eta)x]-c .
\end{equation}
The diagnostic return is positive between $\eta/(p+\eta)$ and
$(1-\eta)/(2-p-\eta)$ and zero outside that open interval.
\end{proposition}
\begin{proof}
For each finding the agent can choose the action with the greater joint probability
of that finding and the corresponding state. Her ex ante probability of being right
after a test is therefore $\max\{xp,(1-x)\eta\}+\max\{x(1-p),(1-x)(1-\eta)\}$. In
the open interval in the statement, a positive and a negative favor different
actions. Subtracting $A(x)$ gives the diagnostic return. The probability of earning
the bounty is $\eta+(p-\eta)x$.
\end{proof}

As in the one-sided model, it is enough to compare the cost with three values of
the gross return. Let $B_\eta=G_\eta+c$ and $K_\eta=p+\eta+V(p-\eta)$. Write
$c_M^\eta$, $c_B^\eta$, and $c_T$ for $B_\eta$ at one half, the upper boundary of
the diagnostic interval, and one, respectively. The last value is still $c_T=pV$.
\begin{theorem}\label{thm:noisy-geometry}
The noisy testing region has at most two components and is nonempty
exactly when $c<\max\{c_M^\eta,c_T\}$. Two components arise exactly when
\begin{equation}\label{eq:noisy-disconnected-condition}
 K_\eta<2,\qquad c_B^\eta\leq c<\min\{c_M^\eta,c_T\}.
\end{equation}
When both components exist, $c=c_B^\eta$ leaves a single indifferent
posterior between them.
\end{theorem}
\begin{proof}
Up to $1/2$ the gross benefit is weakly increasing. Between $1/2$ and
$(1-\eta)/(2-p-\eta)$ its slope is $K_\eta-2$, after which it is again weakly
increasing. Comparing the returns at the endpoints as in Section~\ref{sec:motives}
proves the claim. The explicit values are derived in \ref{app:noisy-cutoffs}.
\end{proof}

\begin{corollary}\label{cor:false-positive-persistence}
Suppose $\kappa<2$ and $c_B<c<\min\{c_M,c_T\}$ for the one-sided test.
For all sufficiently small $\eta>0$, a nondegenerate gap still
separates the two testing components.
\end{corollary}
\begin{proof}
At $\eta=0$ we have $K_\eta<2$ and $c_B^\eta<c<\min\{c_M^\eta,c_T\}$. In
particular, even the weak inequality in \eqref{eq:noisy-disconnected-condition}
holds strictly. The critical returns and $K_\eta$ depend continuously on $\eta$, so
all these inequalities persist for small positive $\eta$.
\end{proof}

If we let $\eta=p$, the finding has the same distribution in both states. Every
type then faces the constant net gain $pV-c$ and makes the same testing decision.

Return to $0<\eta<p<1$ and allow the next agent to investigate after either
finding. Once the public has seen the attempt, a positive or a negative finding
multiplies the resulting odds by
\begin{equation}\label{eq:noisy-outcome-odds}
 \frac{p}{\eta}\quad\text{on a positive finding},\qquad
 \frac{1-p}{1-\eta}\quad\text{on a negative finding}.
\end{equation}
Thus neither state can be excluded at a finite history. To see whether testing can
continue indefinitely, consider the endpoint gains in
\eqref{eq:noisy-gain-decomposition}. They are
\begin{equation}\label{eq:noisy-zero-posterior-gain}
 G_\eta(0)=\eta V-c,\qquad G_\eta(1)=pV-c.
\end{equation}
For $0<c\leq\eta V$ the expected payment alone strictly exceeds the cost at every
interior belief, including at $c=\eta V$, because state one still has positive
probability. Every agent then tests, whichever state is true.

For higher costs the answer can depend on the state. In state~0 there are almost
surely only finitely many attempts if $c>\eta V$. In state~1 the same is true if
$c\geq pV$. To see this, suppose that investigators keep testing indefinitely. The
frequency of positives would reveal the state. In state zero this would eventually
make both types sufficiently pessimistic to decline if $c>\eta V$. In state one it
would make both sufficiently optimistic to decline if $c\geq pV$. At $c=pV$ an
agent close enough to certainty has zero diagnostic gain, and her expected payment
is strictly below $pV$ at interior beliefs. The proof in
\ref{app:stopping-extensions} takes account of the fact that agents choose when to
investigate.

The discrepancy with Theorem~\ref{thm:finite-investigation} can occur even when the
false-positive rate is arbitrarily small. Keep $p=\pi=1/2$, $q=3/4$, $V=2$, and
$c=7/10$ as in Example~\ref{ex:disconnected-region} and \ref{app:submitted-exits}.
For every $0<\eta<1/14$ the gap in private beliefs has positive length and
\begin{equation}\label{eq:noisy-infinite-tests}
 \Pr_1(N_I=\infty)\ \geq\ \frac38-\frac{7\eta}{4}\ >\ \frac14 .
\end{equation}
A first positive can make the public so optimistic that both types test for the
bounty and choose one whatever they find. In state one there is a positive
probability that public beliefs stay in this region forever. The bound in
\ref{app:stopping-extensions} keeps the other parameters fixed as $\eta$ tends to
zero. If the sponsor were to close testing after the first positive, the chance of
a positive at each attempt would again give a geometric bound on their number in
either state. In the noisy model a sponsor following that rule could close the
investigation while both states were still possible.

\subsection{Privately heterogeneous investigation costs}
\label{subsec:heterogeneous-costs}

Now return to conclusive success and let each agent privately observe the cost
$C_t$ of investigating, after she receives her signal. Costs are independent across
agents and independent of the state, signals, and test draws. Their continuous cdf
$F$ has full support on $[\underline c,\overline c]$, with
$0<\underline c<\overline c\leq\infty$. Denote the gross return to the agent by
$B(x)=xpV+D_p(x)$.

\begin{proposition}\label{prop:heterogeneous-costs}
The testing probability and its positive-probability region are
\[
 \iota(x)=F(B(x)),\qquad
 \{x:\iota(x)>0\}=\mathcal X_I(p,\underline c,V).
\]
Consequently \eqref{eq:disconnected-condition}, evaluated at
$c=\underline c$, determines when testing occurs on two separate
intervals. Where $B$ and $F$ are differentiable and $F'(B(x))>0$,
$\iota'$ has the sign of $B'$.
\end{proposition}
\begin{proof}
At belief $x$ the agent tests for precisely those costs $C_t<B(x)$. Continuity of
$F$ gives the probability in the statement, and full support makes it positive
exactly when $B(x)>\underline c$. Differentiating the composition gives the last
assertion.
\end{proof}

Let $h_s=F(B(x^s(\mu)))$ be the probability that type $s$ investigates. If an
attempt has positive probability, observing it multiplies the public odds by
\begin{equation}\label{eq:cost-mixed-choice}
 \frac{qh_1+(1-q)h_0}{(1-q)h_1+qh_0}.
\end{equation}
For a feasible abstention the same calculation uses $1-h_s$ in place of $h_s$. The
choice therefore identifies a signal if only one type makes that choice with
positive probability. When both types make it at different positive rates the
observer learns something about their signal without identifying it. At equal
positive rates she learns nothing from the choice. After an attempt, a failure
still multiplies odds by $1-p$ before the action is observed.

\begin{corollary}\label{cor:heterogeneous-finite-investigation}
With $\underline c>0$ the number of attempts is finite almost surely
in either state. The state-one bounds in \eqref{eq:attempt-tail-state-one}
continue to hold.
\end{corollary}
\begin{proof}
In state one, for an agent to make the $n$th attempt the preceding $n-1$ must have
failed, giving the same bound as before. In state zero, on the event of infinitely
many investigations the public belief would converge to zero by the posterior
argument used to prove Theorem~\ref{thm:finite-investigation}. Eventually both
private beliefs would satisfy $B(x)=\kappa x<\underline c$. Even the lowest
possible cost would then exceed the gross return. Both types would abstain and take
action zero forever, which contradicts infinitely many investigations.
\end{proof}

\begin{remark}\label{rem:zero-cost-atom}
If costs instead have an atom at zero, $p<1$, and $V>0$, every zero-cost
agent tests at every interior history because $xpV>0$. In state~0
success never occurs, all finite-history beliefs are interior, and
infinitely many zero-cost agents arrive almost surely.
\end{remark}

To see the difference between finitely many investigations and a finite date at
which beliefs stop moving, take $p=1/2$, $V=0$, $q=2/3$, $\pi=1/4$, and uniform
costs on $[1/6,2/3]$. In either state there is positive probability that no agent
ever tests. On this path the public nevertheless revises its belief downwards after
each abstention. At any finite date the next agent with a high signal may still
draw a cost low enough to investigate. The public approaches the lower no-test
boundary without reaching it. Thus the bound on attempts in
Corollary~\ref{cor:heterogeneous-finite-investigation} survives even on histories
for which the finite-date conclusion of Theorem~\ref{thm:terminal-outcomes} fails.
The construction is in \ref{app:stopping-extensions}.

\section{Conclusion}\label{sec:conclusion}

An agent can have good reason to investigate a state about which she is already
willing to act. A bounty gives her a return from establishing the state for others.
What her successors learn depends in part on what they would have learned from her
action alone. If that action already discloses her signal, seeing her start the
test can disclose the same news a little earlier. If the action conceals the
signal, the willingness to test may reveal something new. In either case a failure
lowers the belief reached after observing the attempt.

Consider what happens to an optimistic consensus. A bounty can make testing
worthwhile for the favorable type at every belief at which an optimistic cascade
could otherwise begin. In state zero this can eliminate the wrong cascade. In state
one the consensus would have prescribed the correct action. Some of the additional
investigations establish the truth, but on other histories their failures lead
successors to abandon the correct action. This is how a larger bounty can make both
proof and error more likely in the same state. A reward that declines when proof is
already expected can let the optimistic cascade survive. With a fixed reward, an
interval of private beliefs at which agents decline can also sustain it, provided
both signals lead into that interval at a reachable public belief.

Our comparison concerns the action on which agents eventually settle. A sponsor
choosing what bounty to offer would also need to value the actions taken along the
way and account for the cost of producing the evidence. Another consideration is
what counts as a finding. If even a mistaken positive earns a new payment, agents
may continue to investigate indefinitely in state one, however small the
false-positive rate in our example. With conclusive success and a private cost
bounded away from zero, attempts end almost surely. Yet observing an agent pass up
the opportunity to test can continue to change the public belief at every finite
date.

% =============================================================================
% Appendices
% =============================================================================

\clearpage
\appendix
% Keep the journal copy's appendix references and equation numbers.
\renewcommand{\thesection}{Appendix~\Alph{section}}
\renewcommand{\theequation}{\Alph{section}.\arabic{equation}}

\section{Auxiliary derivations and boundary cases}\label{app:proofs}
\setcounter{table}{0}
\renewcommand{\theHtable}{appendix.\Alph{section}.\arabic{table}}
\renewcommand{\thetable}{\Alph{section}.\arabic{table}}
\renewcommand{\theproposition}{\Alph{section}.\arabic{proposition}}
\renewcommand{\thelemma}{\theproposition}

We collect the cutoff calculations, the boundary cases, and the proofs comparing
the two states. \hyperref[app:calculations]{Online Appendix~B} gives the complete records in the finite examples
and bounds the two probabilities reported as decimals in
Table~\ref{tab:leading-errors}.

\subsection{Posterior maps}\label{app:posterior-maps}

To find the public belief at which an agent with signal $s$ reaches a given private
belief $z$, invert the signal updates. This gives
\begin{equation}\label{eq:inverse-signal-maps}
\begin{aligned}
 \psi_1(z)
 &\equiv
 \frac{z(1-q)}{z(1-q)+(1-z)q},\\
 \psi_0(z)
 &\equiv
 \frac{zq}{zq+(1-z)(1-q)},
\end{aligned}
\qquad
 x^s(\psi_s(z))=z.
\end{equation}

\begin{lemma}\label{lem:appendix-posterior-identities}
Let $\mu,z\in(0,1)$, $q\in(1/2,1)$, and $p\in(0,1]$. The signal maps in \eqref{eq:private-posteriors} satisfy
\begin{align}
 O\!\left(x^1(\mu)\right)
 &=\Lambda O(\mu),
 &
 O\!\left(x^0(\mu)\right)
 &=\Lambda^{-1}O(\mu),
 \label{eq:app-signal-odds}\\
 x^1(\mu)-\mu
 &=\frac{\mu(1-\mu)(2q-1)}
 {\mu q+(1-\mu)(1-q)}>0,
 \label{eq:app-positive-signal-difference}\\
 \mu-x^0(\mu)
 &=\frac{\mu(1-\mu)(2q-1)}
 {\mu(1-q)+(1-\mu)q}>0.
 \label{eq:app-negative-signal-difference}
\end{align}
Both maps are strictly increasing. Their inverses are the functions $\psi_s$ in \eqref{eq:inverse-signal-maps}, and
\begin{equation}\label{eq:app-inverse-order}
 \psi_0(z)-\psi_1(z)
 =
 \frac{z(1-z)(2q-1)}
 {\bigl[zq+(1-z)(1-q)\bigr]
 \bigl[z(1-q)+(1-z)q\bigr]}
 >0.
\end{equation}

Whenever failure has positive probability, the failure map satisfies
\begin{align}
 1-\Phi_p(x)
 &=\frac{1-x}{1-px},
 &
 O\!\left(\Phi_p(x)\right)
 &=(1-p)O(x),
 \label{eq:app-failure-identities}\\
 \Phi_p(x)-\frac12
 &=\frac{(2-p)x-1}{2(1-px)}.
 \label{eq:app-failure-half}
\end{align}
Thus $\Phi_p(x)\geq1/2$ if and only if
$x\geq\bar x_p=1/(2-p)$. The sole point at which the denominator $1-px$ vanishes is $(p,x)=(1,1)$, where failure is impossible.
\end{lemma}

\begin{proof}
Dividing each posterior in \eqref{eq:private-posteriors} by its complement gives
\eqref{eq:app-signal-odds}. For each signal its odds are a positive constant times
the public odds, so each signal map is strictly increasing. Subtracting the prior
gives \eqref{eq:app-positive-signal-difference} and
\eqref{eq:app-negative-signal-difference}. Solving $x^s(\mu)=z$ for $\mu$ yields
$\psi_s(z)$, and taking the difference of these two expressions gives
\eqref{eq:app-inverse-order}. For the failure identities
\eqref{eq:app-failure-identities} and \eqref{eq:app-failure-half}, substitute from
\eqref{eq:failure-map}. Whenever the outcome is feasible, its probability $1-px$ is
positive, so multiplying by this denominator preserves the inequality.
\end{proof}

Suppose that the private testing set omits just the belief $z$ between its two
components. For both types to abstain in an upper cascade the public belief would
have to be at least $\psi_0(z)$ for the low type and at most $\psi_1(z)$ for the
high type. By \eqref{eq:app-inverse-order} the first bound exceeds the second.

\subsection{Cutoffs for the private testing decision}
\label{app:gain-boundaries}
On substituting the diagnostic value into
Proposition~\ref{prop:gain-decomposition}, the gain from investigation becomes
\begin{equation}\label{eq:gain-piecewise}
G(x)=
\begin{cases}
 \kappa x-c,&0\leq x\leq1/2,\\
 1-c+(\kappa-2)x,&1/2\leq x\leq\bar x_p,\\
 Wx-c,&\bar x_p\leq x\leq1 .
\end{cases}
\end{equation}
The gain starts at $-c$. At the next three endpoints $1/2$, $\bar x_p$, and one it
takes the values $c_M-c$, $c_B-c$, and $c_T-c$. The relative sizes of these
critical costs follow from
\[
 c_M-c_B=\frac{p(2-\kappa)}{2(2-p)},\qquad
 c_T-c_B=\frac{W(1-p)}{2-p}.
\]
For Theorem~\ref{thm:investigation-geometry} we distinguish three cases according
to the slope of the middle piece. In each case an interior root is excluded from
the testing set, since an indifferent agent abstains.

\noindent\emph{(i) A declining middle piece ($\kappa<2$).}
Use the roots $\ell,u,h$ in \eqref{eq:declining-region-roots}. We need $h$ only
when the upper piece is profitable.
\begin{equation}\label{eq:geometry-kappa-below-two}
\mathcal X_I=
\begin{cases}
 (\ell,1],
 &0<c<c_B,\\[0.35em]
 (\ell,\bar x_p)\cup(\bar x_p,1],
 &c=c_B<c_T,\\[0.35em]
 (\ell,u)\cup(h,1],
 &c_B<c<c_M\ \text{and}\ c<c_T,\\[0.35em]
 (\ell,u),
 &c<c_M\ \text{and}\ c\geq c_T,\\[0.35em]
 (h,1],
 &c\geq c_M\ \text{and}\ c<c_T,\\[0.35em]
 \varnothing,
 &c\geq c_M\ \text{and}\ c\geq c_T.
\end{cases}
\end{equation}

\smallskip
\noindent\emph{(ii) A flat middle piece ($\kappa=2$).}
With a flat middle piece, $c_M=c_B=1$ and $c_T=W=2-p$. The testing set is
\begin{equation}\label{eq:geometry-kappa-equal-two}
\mathcal X_I=
\begin{cases}
 (c/2,1],&0<c<1,\\[0.35em]
 (c/W,1],&1\leq c<W,\\[0.35em]
 \varnothing,&c\geq W.
\end{cases}
\end{equation}
For $p=1$ we have $W=1$, so the cost interval in the middle line is empty.

\smallskip
\noindent\emph{(iii) A rising middle piece ($\kappa>2$).}
The gain now increases across the middle piece as well. When $c\geq c_T$ no type
tests. For $c<c_T$ the testing set is
\begin{equation}\label{eq:geometry-kappa-above-two}
 \mathcal X_I=(\tau(c),1],
\end{equation}
where
\begin{equation}\label{eq:monotone-threshold}
 \tau(c)=
 \begin{cases}
 c/\kappa,&0<c\leq c_M,\\[0.45em]
 (c-1)/(\kappa-2),&c_M<c\leq c_B,\\[0.45em]
 c/W,&c_B<c<c_T.
 \end{cases}
\end{equation}
Here we keep only the lines satisfying $c<c_T$. At $c=c_M$ and $c=c_B$ the adjacent
formulas give the same cutoff.

The signs at the three endpoints determine the intervals just listed. When the
middle piece declines and there are two components, the distance between their
adjacent roots is
\begin{equation}\label{eq:app-gap-length}
 h-u=\frac{(2-p)(c-c_B)}{W(2-\kappa)}.
\end{equation}
At $c=c_B<c_T$ the agent abstains at a single belief between the two components. If
both components remain and $c>c_B$, the interval between them has positive length.

\subsection{Likelihood of a complete public record}
\label{app:record-likelihood}

Consider an observer at public history $H$. Let $\sigma^I(i\mid H,s)$ be the
probability that type $s$ makes choice $i$. In state $\theta$ the observer assigns
that choice probability
\begin{equation}\label{eq:investigation-selection}
 \alpha^\theta(i\mid H)=\sum_{s=0}^1 \Pr(s\mid\theta)\,\sigma^I(i\mid H,s).
\end{equation}
For a feasible $i$, observing the choice multiplies the odds by
$\alpha^1(i\mid H)/\alpha^0(i\mid H)$. If the agent investigates and fails, the
odds at the attempt are then multiplied by $1-p$. At $p<1$ her ensuing action
contributes its likelihood ratio conditional on that choice and outcome. At $p=1$
the failure already establishes state zero. The same sequence of updates applies
with private costs, even though the observer who knows the signal cannot predict
the investigation decision with certainty in general.

For the deterministic testing rule we can obtain the update in one step. Fix an
interior public belief $\mu$ and a feasible record $r$ without success. Let
$\mathcal S(r,\mu)\subseteq\{0,1\}$ contain the types whose decision and action
generate $r$. Put $f(r)=1$ for a failed investigation and $f(r)=0$ for abstention.
The likelihood ratio of the observed record is then
\begin{equation}\label{eq:app-record-likelihood-ratio}
 \frac{\Pr(r\mid\theta=1,\mu)}
 {\Pr(r\mid\theta=0,\mu)}
 =
 (1-p)^{f(r)}
 \frac{\displaystyle
 \sum_{s\in\mathcal S(r,\mu)}\Pr(s\mid\theta=1)}
 {\displaystyle
 \sum_{s\in\mathcal S(r,\mu)}\Pr(s\mid\theta=0)}.
\end{equation}
The second factor reduces to $\Lambda^{2s-1}$ when the observer can infer $s$.
Summing over both signals gives one when she cannot distinguish them. The four
possible updates are therefore
\begin{equation}\label{eq:app-record-posterior-odds}
 \frac{O(\mu'(r))}{O(\mu)}
 =
 \begin{cases}
 \Lambda^{2s-1},
 &r\text{ reveals }s\text{ and contains no investigation},\\
 1,
 &r\text{ pools signals and contains no investigation},\\
 (1-p)\Lambda^{2s-1},
 &r\text{ reveals }s\text{ and contains a failure},\\
 1-p,
 &r\text{ pools signals and contains a failure}.
 \end{cases}
\end{equation}
For an abstention the outcome factor in \eqref{eq:app-record-likelihood-ratio} is
one, also at $p=1$. The resulting expressions give the last four lines of
\eqref{eq:exact-public-transition}. A success gives the first line, since its
state-zero probability is zero.

In particular, when the attempt has already disclosed the signal, the action after
failure adds no further signal information. This is the same inference obtained by
updating after each observation in Section~\ref{subsec:belief-transitions}.

\subsection{The posterior-martingale step}
\label{app:martingale-step}

For the argument in Theorem~\ref{thm:finite-investigation}, let
$\mathcal H_\infty=\sigma(\bigcup_t\mathcal H_t)$ be the information in the entire
public history. As before, once the state is revealed its posterior remains fixed.

\begin{lemma}\label{lem:appendix-null-event}
Let $E\in\mathcal H_\infty$ be a public-history event such that $\Pr(E\mid\theta=1)=0$. Then the limiting public posterior is zero almost surely on $E$.
\end{lemma}

\begin{proof}
Under the prior distribution, the public posterior is the bounded martingale
\[
 \mu_t
 =
 \mathbb E\!\left[\mathbf 1\{\theta=1\}\mid\mathcal H_t\right].
\]
The convergence theorem for conditional expectations on an increasing filtration
gives
\[
 \mu_t\longrightarrow
 \mu_\infty
 \equiv
 \mathbb E\!\left[\mathbf 1\{\theta=1\}\mid\mathcal H_\infty\right]
\]
almost surely and in $L^1$. We may therefore integrate over the public event
$E\in\mathcal H_\infty$ to obtain
\[
 \mathbb E[\mathbf 1_E\mu_\infty]
 =
 \Pr(E\cap\{\theta=1\})
 =
 \pi\Pr(E\mid\theta=1)
 =0.
\]
Since $\mathbf 1_E\mu_\infty$ is nonnegative and has expectation zero, it vanishes
almost surely.
\end{proof}

Apply the lemma to the event $E$ of infinitely many investigations. Conditional on
state~1, assign an independent Bernoulli draw with success probability $p$ to each
potential attempt, and reveal the draw only after the agent chooses to make that
attempt. If an $n$th attempt is made, the first $n-1$ draws must all have been
failures. It follows that
\[
 \Pr(E\mid\theta=1)
 \leq
 \lim_{n\to\infty}(1-p)^{n-1}
 =0.
\]
Consequently $\mu_t\to0$ almost surely on $E$ by
Lemma~\ref{lem:appendix-null-event}. To see why this rules out $E$ in state zero as
well, observe that for
\[
 \varepsilon_0
 =
 \min\left\{\frac14,\frac{c}{2\kappa}\right\},
\]
every $x\leq\varepsilon_0$ lies on the first piece of the gain and satisfies
\begin{equation}\label{eq:app-safe-gain}
 G(x)=\kappa x-c\leq-\frac c2<0.
\end{equation}
Once $0<\mu\leq\delta_0=\psi_1(\varepsilon_0)$, the two private posteriors satisfy
$x^0(\mu)<x^1(\mu)\leq\varepsilon_0<1/2$. Both types decline and produce the same
action-zero record. The public belief then stays fixed. Thus infinitely many
investigations would themselves lead agents into a region where none can occur.
This proves the state-zero claim in Theorem~\ref{thm:finite-investigation} for
either shape of testing set.

\subsection{Cascade-band and upper-boundary checks}
\label{app:trap-boundaries}

Using \eqref{eq:private-posteriors} and action one at indifference gives
\begin{align}
 x^1(\mu)<\frac12
 &\quad\Longleftrightarrow\quad \mu<1-q,
 \label{eq:app-lower-action-band}\\
 x^0(\mu)\geq\frac12
 &\quad\Longleftrightarrow\quad \mu\geq q.
 \label{eq:app-upper-action-band}
\end{align}
At $\mu=1-q$ the high type is just indifferent and takes one, so that belief is
excluded from the lower band. At $\mu=q$ it is the low type who is indifferent and
takes one, so that belief belongs to the upper band.

In the lower band $x^0(\mu)<x^1(\mu)$ and both types have gain $G(x)=\kappa x-c$.
It therefore suffices to check $x^1(\mu)\leq c/\kappa$. Inverting this inequality
gives \eqref{eq:lower-trap-explicit}. When $c<c_M$ the high type abstains at the
right endpoint.

For a connected upper testing interval $(r,1]$, the high type must still decline at
$\mu\geq q$. This gives $[q,\psi_1(r)]$ if the right endpoint is at least $q$, and
an empty set otherwise. For a bounded central testing interval $(\ell,u)$ it is the
low type who must reach $u$, giving $[\psi_0(u),1)$, empty if $u=1$. For a split
testing set the two bounds are those in \eqref{eq:upper-trap-gap}. When they
coincide there is one candidate belief. It contributes to an upper cascade from a
specified prior only if a finite history reaches it.

Suppose that $p<1$ and let $\mu^F$ be public belief after a failed investigation
and the resulting action. For $c<c_M$ write $\bar\mu_0\equiv\psi_1(c/\kappa)$. For
either kind of failed record, the following inequality is necessary and sufficient
for lower entry.
\begin{equation}\label{eq:failure-lower-trap-odds}
 \begin{cases}
 (1-p)O(\mu)\leq O(\bar\mu_0),
 &\text{for a pooling failure},\\
 (1-p)\Lambda^{2s-1}O(\mu)\leq O(\bar\mu_0),
 &\text{for a failure revealing $s$}.
 \end{cases}
\end{equation}
For $c\geq c_M$ the right-hand threshold becomes $O(1-q)=\Lambda^{-1}$ and the
inequality is strict. For $p=1$, every feasible failure gives $\mu^F=0$ and ends
investigation by revealing $\theta=0$. The interior lower trap $\mathcal T_0$
excludes this conclusive outcome.

One may also ask whether all sufficiently optimistic interior beliefs are upper
traps. The necessary and sufficient condition is
\begin{equation}\label{eq:upper-neighborhood-condition}
 c>pV,
 \quad\text{or}\quad
 c=pV\ \text{ and }\
 \bigl[p<1\ \text{ or }\ V\geq1\bigr].
\end{equation}

To verify \eqref{eq:upper-neighborhood-condition}, consider an imperfect test
first. Since $\bar x_p<1$ for $p<1$, the gain just below one is
\[
 G(x)=pVx-c.
\]
This is nonpositive throughout a left neighborhood exactly when $c\geq pV$. With a
perfect test, $\bar x_p=1$ and the middle piece of the gain extends all the way to
one. It gives
\begin{equation}\label{eq:app-perfect-test-upper-gain}
 G(x)=1-c+(V-1)x,\qquad x\in[1/2,1].
\end{equation}
This gain is negative near one for $c>V$ and positive there for $c<V$. At $c=V$ it
reduces to
\begin{equation}\label{eq:app-perfect-test-equality}
 G(x)=(1-V)(1-x).
\end{equation}
Thus the upper-trap neighborhood exists at equality if and only if $V\geq1$. For
$p=1$, $0<V<1$, and $c=V$, the gain $(1-V)(1-x)$ is strictly positive at every
$x\in[1/2,1)$. The small chance that a perfect test will correct the action
suffices to keep investigation profitable even arbitrarily close to certainty.
\subsection{Entry probabilities under a fixed bounty}\label{app:submitted-exits}
\begin{proof}[Proof of Proposition~\ref{prop:submitted-error-comparison}]
Let $d=1-q$ and $\Lambda=q/d$. Expressed in private odds, the testing set is
$(7/8,3/2)\cup(7/3,\infty)$, and public odds $O\leq7/(8\Lambda)$ constitute the
lower trap. Suppose first that $1<\Lambda<8/7$. Both signal types investigate at
the prior and generate the same record on failure. A failure takes the public odds
to $1/2$, which already belongs to the lower trap. Hence $\Pr_0(L)=1$ and
$\Pr_1(L)=1/2$.

Now take $\Lambda\geq8/7$. An initial low type abstains and chooses zero, revealing
her signal and sending odds to $1/\Lambda$. Another low takes them to
$1/\Lambda^2\leq7/(8\Lambda)$. If the second signal is high instead, the agent's
private odds are one, she tests, and a failure leaves public odds $1/2$. For
$\Lambda\leq7/4$ this is already a lower exit. For $\Lambda>7/4$, give the
following agent a low signal. She abstains and chooses zero. This identifies the
low signal, because at public odds $1/2$ the high type either tests or has private
odds at least one and chooses one. Public odds therefore fall to $1/(2\Lambda)$,
inside the lower trap. Adding the probabilities of these disjoint histories gives
\[
 \Pr_1(L)\ \geq \begin{cases}
 d^2+dq/2,&8/7\leq\Lambda\leq7/4,\\
 d^2(1+q/2),&\Lambda>7/4 .
 \end{cases}
\]
Compare each lower bound with the benchmark $d^2/(1-dq)$. In the middle range the
ratio exceeds one because $(1+q/(2d))(1-dq)>1$ for $q>d$ and $dq\leq1/4$. In the
last range, $q>7/11$ implies $(1+q/2)(1-dq)-1=q(q^2+q-1)/2>0$. In the first range
the comparison is simply $1/2>1/(1+\Lambda+\Lambda^2)$.

For an upper trap to exist, its odds must lie between $3\Lambda/2$ and
$7/(3\Lambda)$. This requires $\Lambda^2\leq14/9$. We have already dealt with
$1<\Lambda<8/7$. For $8/7\leq\Lambda\leq\sqrt{14}/3$, an initial high tests, and a
failure takes odds to $\Lambda/2\leq7/(8\Lambda)$. After an initial low, both
state-zero continuations reach the lower trap as shown above. Thus $\Pr_0(L)=1$
whether the upper trap is empty or simply unreachable. At each lower entry the odds
are at most $7/(8\Lambda)$. Summing over those histories gives
$\Pr_1(L)=\mathbb E_0[O(\mu_{\tau+1})]\leq7/(8\Lambda)$. Without testing, the two
errors sum to $(\Lambda+2)/(1+\Lambda+\Lambda^2)>1/\Lambda$. Weighting the states
equally proves the last claim.
\end{proof}

For the value at $q=3/4$, start with unit mass at odds one in each state's
calculation. At any record before proof the odds are $3^k2^{-j}$, where $k$ is the
number of revealed highs less the number of revealed lows and $j$ counts failed
tests. For each current belief, distribute its mass using
\eqref{eq:absorption-state-zero} or \eqref{eq:absorption-state-one}. Assign proof
and first entries to their respective outcomes, and add together active histories
that give identical odds. All transitions and cutoff comparisons use rational
numbers. After 100 agents the total active mass is below $8\cdot10^{-23}$ in either
state. The lower mass already absorbed in state one, rounded down, and that mass
plus the active mass, rounded up, give the bounds
\[
 0.141489033705 < \Pr_1(L) < 0.141489033706 .
\]
The eventual proof probability in state one is one minus this lower-exit
probability. The bounds yield the decimals in Table~\ref{tab:leading-errors}, while
the preceding finite histories establish the direction of the comparisons.

\begin{proof}[Proof of Theorem~\ref{thm:bounty-exits}]
Suppose first that $c<p(1+V)/2$. At private exponent $j\leq0$ the agent gains $px$
in accuracy from a test. She therefore tests precisely when $\Lambda^{-j}<C$. Let
$a=1-\lceil\log_\Lambda C\rceil\leq0$ be the lowest exponent in this range at which
she tests. In particular, the agent at $j=0$ tests. At $j\geq1$, action one is
optimal even after a failure, so the criterion is $V>t_j=(c/p)(1+\Lambda^{-j})$. At
equality she declines.

Consider first a path into the lower trap. At public exponent $k\leq a-2$ both
types abstain and take zero. At $a-1\leq k\leq-1$ the high type investigates, and
at $k\geq0$ even her action upon abstaining would be one. An inconclusive record
can reduce $k$ by at most two. At $k=a-1$ only the high type investigates. Her
failed test leaves the exponent at $a-1$, and the low type's abstention takes it
down by one. No first entry can therefore fall below $a-2$. Since the lower trap
starts there, every first lower entry is at $a-2=k_L$.

For upper entry, the low type at $k=1$ has exponent zero and tests. If $V\leq t_3$,
at $k=2$ even the high type, whose private exponent is three, declines to test.
Both then take action one. To reach $k=2$, the first two signals must be high. The
corresponding private exponents are one and two, and both agents abstain and reveal
their signal. A low before entry either produces proof or takes the exponent below
zero. At $k=-1$ the high type tests, preventing a return to zero, and no single
record can raise $k$ by more than one. Now suppose $V>t_3$. The high type at $k=2$
also tests. Her failure cannot raise $k$, nor can a low signal there. Thus no
higher exponent and no upper trap can be reached.

It follows that state zero reaches the upper trap with probability $u$ and, by
Theorem~\ref{thm:terminal-outcomes}, the lower trap with probability $1-u$. Since
the initial odds are one, the mass of any first-entry history in state one is its
state-zero mass times the odds at entry. The lower entries all have odds
$\Lambda^{k_L}$ and all possible upper entries have odds $\Lambda^2$. Summation
yields \eqref{eq:closed-state-one}.

Finally take $c\geq p(1+V)/2$, so that every type $j\leq0$ declines. If
$V\leq t_2$, the types at $j=1,2$ also decline. We recover the benchmark with
transient exponents $-1,0$ and absorbing exponents $-2,1$. If $V>t_2$, consider
what happens at $k=1$. The high type tests and her failure leaves the public at
$k=1$. The low type abstains and sends it to zero. Thus histories starting below
one cannot cross $k=1$ or enter an upper trap. At $k=-1$ the two agents abstain and
reveal their respective signals, giving $k=-2$ as the sole lower entry. Applying
the same likelihood identity proves the second part.
\end{proof}

\subsection{Perturbing the test and signal precisions}\label{app:off-grid}
The ordering of the errors in Section~\ref{subsec:absorption-probabilities} also
holds away from the exact matching condition. Keep $\pi=1/2$ and take $V_1=1/10$,
$V_2=1/4$, and $V_3=19/40$. At $(q,p,c)=(2/3,1/2,1/5)$ the state-one errors in
increasing order are $1/9$ at $V_2$, $1/8$ at $V_3$, $1/7$ without a test, and
$2/9$ at $V_1$. We show that the ordering persists on an open set of parameters
with $(1-p)\Lambda>1$.

Fix a finite horizon. At the reference parameters the public exponents reachable
under these bounties run from $-3$ to $2$. The associated private exponents are
$j=-4,\ldots,3$. In that order, the bounties for which each type is indifferent are
\[
 29/5,\quad 13/5,\quad 1,\quad 1/5,\quad -1/5,\quad
 3/5,\quad 1/2,\quad 9/20.
\]
None of the three chosen bounties makes a testing type or an abstaining type
indifferent. There can still be action ties. Put $B=(1-p)\Lambda$. Along a fixed
sequence of records the public odds, and the private odds when choosing an action,
have the form $\Lambda^kB^j$, where $k$ is an integer and $j\geq0$ counts the
failures already included. If the agent was tied at the reference parameters, then
$k=0$. Her odds remain at least one on the $B\geq1$ side and she continues to
choose one. Every strict choice persists locally by continuity. Induction to the
chosen horizon therefore preserves all records and trap entries for sufficiently
small perturbations on this side. Their state-conditional probabilities vary
continuously there.

By Theorem~\ref{thm:terminal-outcomes} we can take the horizon large enough that
the unresolved mass at the reference parameters is arbitrarily small. This mass is
itself the probability of a finite collection of histories and stays close under
the same perturbations. Each eventual exit probability lies between the mass
already absorbed at that exit and that mass plus the unresolved mass. This gives
one-sided continuity. Since the four errors are strictly ordered at the reference
parameters, the same inequalities hold on the open portion with $B>1$, where $q$,
$p$, and $c$ may vary independently.

\subsection{Reward schedules and entry into a testing gap}
\label{app:reward-comparisons}

\begin{proof}[Proof of Proposition~\ref{prop:surprise-exits}]
Let public belief tend to one. The two private posteriors then tend to one and
their diagnostic gains vanish. The reward for success tends to zero by assumption,
so both net gains in \eqref{eq:public-reward-gain} tend to $-c$. In a left
neighborhood of one both types therefore decline and take action one. Near zero
both the diagnostic gain and $px^s(\mu)R(\mu)$ vanish, the latter because $R$ is
bounded. This gives a lower region of abstention as well. From below the upper
region a single record can raise odds by at most a factor $\Lambda$, so reachable
interior odds are bounded. With these bounds the same block of low signals as in
Theorem~\ref{thm:terminal-outcomes} forces absorption.

For the linear payment, write the public odds as $\Lambda^k$. At $k\geq0$ the high
type has private exponent $k+1$. Her diagnostic gain is zero, so she buys a test
exactly when
\[
 A>B_k=(c/p)(1+\Lambda^{-k-1})(1+\Lambda^k).
\]
For $k\geq0$ these thresholds are increasing. Take $A\leq B_0$. The high type
declines at $k=0,1,2$. At $k=1$ the low type has posterior $1/2$ and tests since
$D_p(1/2)=p/2>c$. At $k=2$ even she declines, because her threshold is
$(c/p)(1+\Lambda^{-1})(1+\Lambda^2)>B_0$. Two initial highs thus take the public
into the upper trap at $k=2$. A low before then either produces proof or takes $k$
below zero. From $k=-1$ a high signal induces a test at private odds one,
preventing an inconclusive return to zero. This gives \eqref{eq:taper-upper}. Take
now $A>B_0$. Already at $k=0$ a high type tests and her failure cannot raise the
odds. Neither can a low signal there. Since no inconclusive step raises $k$ by more
than one, a path from the prior cannot reach an upper trap.
\end{proof}

\emph{The lower exit under a declining reward.}
Under the hypotheses of \eqref{eq:taper-upper}, denote the upper probability in
state zero by $u_A=\Pr_0(U)$ and the public odds by $t=\Lambda^k$. To locate lower
entry, let $z$ be the positive root of
\begin{equation}\label{eq:taper-root}
 \Lambda(p-c)z^2+
 [p\Lambda(1+A)-c(1+\Lambda)]z-c=0,
 \qquad K=\lfloor\log_\Lambda z\rfloor .
\end{equation}
Then the other exit probabilities are
\[
 \Pr_0(L)=1-u_A,\qquad \Pr_1(L)=\Lambda^K(1-u_A).
\]
The complement of the two cascade probabilities in state one is the proof
probability. To derive the lower entries, observe that at $k\leq-1$ both private
beliefs are at most $1/2$. The high type has the larger gross return from testing.
It is
\[
 g(t)=\frac{p\Lambda t}{1+\Lambda t}
       +\frac{pA\Lambda t}{(1+\Lambda t)(1+t)} .
\]
The first term is strictly increasing. For the second term, the derivative of
$t/[(1+\Lambda t)(1+t)]$ has the sign of $1-\Lambda t^2$, positive on
$0<t\leq1/\Lambda$. We have $g(0)=0$ and $g(1/\Lambda)\geq p/2>c$. Clearing
denominators in $g(z)=c$ gives \eqref{eq:taper-root}. Its positive leading
coefficient and negative constant term give exactly one positive root. Hence
$K\leq-2$. At $k\leq K$ the two types decline and choose zero, while for
$K<k\leq-1$ the high type tests. At $k\geq0$ her abstention would give action one.

It remains to check whether the public can jump across $K$ when it first enters the
lower trap. The low type at $K+1$ has lower private odds and faces a weakly smaller
payment than the high type at $K$, who declines. Thus only the high type tests at
$K+1$. Her failure leaves the exponent there, and the low type's abstention gives
$K$. From any $k\geq K+2$ one record can reduce the exponent by at most two. Hence
each first lower entry has odds $\Lambda^K$. Weighting its state-zero mass by those
odds gives its state-one mass, as claimed.

\smallskip\noindent\emph{The first decision affected by the payment.}
In Example~\ref{ex:two-motives}, take $q=2/3$ and $\pi=1/2$. Until odds first reach
four, fixed $V=1/2$ and declining $R(\mu)=A(1-\mu)$ for either $A=1/2$ or $A=1$
give the transitions below. Each pair lists the next public exponent after a low
and a high signal, in that order. A star marks an agent who tests, so that the
displayed transition occurs on failure.
\[
\begin{array}{c|rrrrr}
 k&-3&-2&-1&0&1\\ \hline
 (k'_0,k'_1)&L&(-3,-2^*)&(-2,-1^*)&(-2^*,1)&(-1^*,2)
\end{array}
\]
With the fixed payment the pair at $k=2$ is $(1,2^*)$. Both declining schedules
yield $U$ there. At the original cost $c=1/5$ the high type at $k=1$ is just
indifferent under the fixed payment and declines. To find the range giving the same
decisions, include every reachable belief, even the absorbing ones. The largest
gross gain of any abstaining type and the smallest gross gain of any testing type
are $1/5$ and $2/9$ for fixed $V=1/2$. Equation~\eqref{eq:public-reward-gain} gives
$13/90$ and $5/24$ for $A=1/2$, and $17/90$ and $1/4$ for $A=1$. Their common cost
range is $[1/5,5/24)$. Throughout its interior the comparisons are strict. At
$5/24$ a type who previously tested under $A=1/2$ abstains. At odds four, where the
rules first differ, the private posteriors are $2/3$ and $8/9$ and neither type
gains in accuracy from a test.

\smallskip\noindent\emph{Width of the gap and changes in cost.}
Keep $p=1/2$, $c=1/5$, and $q=4/7$, and vary the fixed bounty. For $2/5<V<3/5$ the
testing set has two components. Theorem~\ref{thm:investigation-geometry} gives
$8/(7-5V)$ and $2/(5V-2)$ as the two ends of the private gap in odds. The condition
for an upper trap in \eqref{eq:gap-in-odds} becomes
\[
 (4/3)^2\frac{8}{7-5V}\leq\frac{2}{5V-2},
 \quad\text{or equivalently}\quad V\leq191/365 .
\]
At $V=1/2$ the upper-trap odds lie in $[64/27,3]$, which includes $8/3$. At
$V=8/15$ a gap of positive length remains in private beliefs, yet the upper-trap
set is empty. In both examples the first high type abstains and moves public odds
from $2$ to $8/3$. What differs is the decision faced by the next pair of signal
types.

For the cost comparison hold $V=1/2$ fixed. At public odds two the low and high
types' gross returns are $1/4$ and $2/11$, respectively. At odds $8/3$ they are
$1/6$ and $8/41$. Thus for $8/41\leq c<1/4$ the initial decision separates and the
high type's abstention starts the upper cascade. The lower endpoint belongs to this
interval because at $c=8/41$ the high type at odds $8/3$ is indifferent and
declines. At $c=1/4$ even the low type abstains at the prior.

For costs in $[8/41,1/4)$ the lower private testing cutoff is $4c/3$. The pair
$V'=1$, $c'=4c/3$ preserves it and fills the gap since $c'<1/3$. After a first low
and failure the odds still contract as in Proposition~\ref{prop:gap-filling}. The
probability of upper entry in state~0 therefore falls from $3/7$ to zero throughout
that interval. Other decisions can change within it. At odds $9/32$ the high type
has posterior $3/11$ and gross benefit $9/44$ under the original bounty. She tests
at $c=1/5$ and abstains at $c=9/44$.

\subsection{Stopping with noisy outcomes and private costs}
\label{app:stopping-extensions}

\smallskip\noindent\emph{The private cutoffs for a noisy finding.}\phantomsection\label{app:noisy-cutoffs}
Let an agent's belief before the test be $x$. Her beliefs after a positive and a
negative finding are given by
\begin{equation}\label{eq:noisy-outcome-posteriors}
\begin{aligned}
 y^+_\eta(x)
 &=
 \frac{xp}{xp+(1-x)\eta},\\
 y^-_\eta(x)
 &=
 \frac{x(1-p)}
 {x(1-p)+(1-x)(1-\eta)},
\end{aligned}
\end{equation}
in each case provided the finding has positive probability. The corresponding
cutoffs for taking action one are
\begin{equation}\label{eq:noisy-action-thresholds}
 \underline x_\eta
 \equiv\frac{\eta}{p+\eta},
 \qquad
 \overline x_\eta
 \equiv\frac{1-\eta}{2-p-\eta}.
\end{equation}
Since $p>\eta$, the two cutoffs satisfy
\[
 0\leq\underline x_\eta<\frac12
 <\overline x_\eta\leq1.
\]
Thus, following a feasible positive finding, the agent takes action one if and only
if $x\geq\underline x_\eta$. Following a feasible negative finding she does so if
and only if $x\geq\overline x_\eta$.

To calculate the value of the information, fix a finding and compare the joint
probabilities of that finding with state zero and with state one. The larger
probability is its contribution to the expected payoff from the action. Adding the
contributions of the two findings and subtracting the payoff $A(x)$ available
without a test gives
\begin{equation}\label{eq:noisy-diagnostic-value}
 D_{p,\eta}(x)
 =
 \begin{cases}
 0,
 &0\leq x\leq\underline x_\eta,\\[0.4em]
 (p+\eta)x-\eta,
 &\underline x_\eta\leq x\leq1/2,\\[0.4em]
 1-\eta-(2-p-\eta)x,
 &1/2\leq x\leq\overline x_\eta,\\[0.4em]
 0,
 &\overline x_\eta\leq x\leq1.
 \end{cases}
\end{equation}

To see how false positives affect the return from testing, let
\begin{equation}\label{eq:noisy-slope-parameters}
 W_\eta\equiv V(p-\eta),
 \qquad
 K_\eta\equiv p+\eta+V(p-\eta),
\end{equation}
and let $B_\eta(x)=G_\eta(x)+c$ be the benefit before paying for the test. On each
of its four affine pieces the slope is
\begin{equation}\label{eq:noisy-gain-slopes}
 B_\eta'(x)
 =
 \begin{cases}
 W_\eta,
 &0<x<\underline x_\eta,\\
 K_\eta,
 &\underline x_\eta<x<1/2,\\
 K_\eta-2,
 &1/2<x<\overline x_\eta,\\
 W_\eta,
 &\overline x_\eta<x<1.
 \end{cases}
\end{equation}
The benefit therefore rises weakly up to $1/2$ and again above $\overline x_\eta$.
Only between these beliefs can it fall. Its values at the three beliefs that
determine whether the testing region splits are
\begin{equation}\label{eq:noisy-critical-costs}
\begin{aligned}
 c_M^\eta
 &\equiv
 B_\eta\!\left(\frac12\right)
 =
 \frac{p(1+V)+\eta(V-1)}{2},\\
 c_B^\eta
 &\equiv
 B_\eta(\overline x_\eta)
 =
 \frac{V(p+\eta-2p\eta)}{2-p-\eta},\\
 c_T
 &\equiv B_\eta(1)=pV.
\end{aligned}
\end{equation}
Subtracting the value at $\overline x_\eta$ from each of the other two gives
\begin{equation}\label{eq:noisy-critical-ordering}
\begin{aligned}
 c_M^\eta-c_B^\eta
 &=
 \frac{(p-\eta)(2-K_\eta)}
 {2(2-p-\eta)},\\
 c_T-c_B^\eta
 &=
 \frac{V(1-p)(p-\eta)}
 {2-p-\eta}
 \geq0.
\end{aligned}
\end{equation}

Together with continuity, these comparisons and the middle slope yield
Theorem~\ref{thm:noisy-geometry}. In particular, an agent at the bridge declines to
test when her benefit just covers its cost.

\smallskip\noindent\emph{Continued testing after either noisy outcome.}
Take $0<\eta<p<1$. Imagine drawing a sequence of potential findings, one for each
successive attempt. The draws are independent conditional on the state and an agent
decides whether to test before she can see her draw. If agents test infinitely
often, the public eventually observes every member of this sequence. The strong law
then puts the limiting fraction of positives at $\eta$ in state zero and at $p$ in
state one, almost surely. Infinitely many observed tests with limiting frequency
$\eta$ define a public event of probability zero in state one.
Lemma~\ref{lem:appendix-null-event} gives $\mu_t\to0$ on that event, almost surely.
Interchanging the states gives $\mu_t\to1$ on the corresponding event with
frequency $p$.

Now consider the private incentives along these paths. As $\mu_t$ tends to either
endpoint, the private beliefs of both types approach the same endpoint. If
$c>\eta V$, then $G_\eta(0)<0$ and neither type tests sufficiently near zero. If
$c\geq pV$, there is an upper neighborhood where testing has no diagnostic value
and the expected payment net of cost is $V[\eta+(p-\eta)x]-c\leq0$. In either
neighborhood both types abstain and take the same action. Once the public enters,
no subsequent agent tests. This proves the respective state-zero and state-one
finiteness claims in Section~\ref{subsec:noisy-verification}.

For the example in \eqref{eq:noisy-infinite-tests}, substitution gives
$c_B^\eta=1/(3/2-\eta)$, $c_M^\eta=3/4+\eta/2$, and $K_\eta=3/2-\eta$. Whenever
$0<\eta<1/14$, we have $c_B^\eta<7/10<\min\{c_M^\eta,1\}$. At the prior the
expected payment alone is at least $3/4$ for the high type. Even counting the gain
in accuracy, the low type's benefit is only $3/8+3\eta/4<7/10$. A first agent who
tests therefore reveals a high signal. In state one this signal followed by a
positive finding has probability $3/8$ and takes public odds to $O_*=3/(2\eta)>7$.

Above odds seven, even the low type has private belief above $7/10$. Both types
want the expected payment and both take action one even after a negative finding.
As long as the public stays in this region, only the test result changes the odds.
Let $O_n$ be the odds after $n$ further tests, starting at $O_0=O_*$, and let $T$
be their first entry at or below seven. Under state one the reciprocal odds
multipliers $\eta/p$ and $(1-\eta)/(1-p)$ have mean one. Hence $O_*/O_{n\wedge T}$
is a nonnegative martingale with initial value one. On $\{T\leq n\}$ it is at least
$O_*/7$, which gives $\Pr_1(T\leq n\mid O_*)\leq7/O_*$ at every $n$. With
probability at least $1-7/O_*$ conditional on reaching $O_*$, the public stays
above seven forever and every subsequent agent tests. Multiplication by $3/8$ gives
the lower bound $3/8-7\eta/4$ in \eqref{eq:noisy-infinite-tests}.

\smallskip\noindent\emph{Finite attempts with continued learning from abstention.}
Take the one-sided test again, with $p=1/2$, $V=0$, $q=2/3$, and $\pi=1/4$, and let
costs be uniform on $[1/6,2/3]$. The prior odds are $1/3$. At public odds
$1/4<O\leq1/3$ even the favorable type's private belief is below $1/2$. Both types
would choose zero without a test, and the gross benefit of testing is $x/2$. This
benefit never covers the low type's cost. The high type's probability of
investigating is
\[
 h(O)=\frac{2O}{1+2O}-\frac13.
\]
An abstention followed by action zero is thus more likely in state zero. The new
public odds $T(O)$ satisfy
\[
 T(O)=O\,\frac{1-(2/3)h(O)}{1-(1/3)h(O)},\qquad
 T(O)-\frac14=\left(O-\frac14\right)\frac{5(1+O)}{5+7O}.
\]
On $[1/4,1/3]$ the multiplier of $O-1/4$ is positive and at most $25/27$. Along a
history of abstentions the odds stay strictly above $1/4$ at every finite date and
converge to it at least geometrically fast. Since $h(O)=4(O-1/4)/[3(1+2O)]$, the
sum of the testing probabilities along this history is finite. Write $b_\theta$ for
the probability of a high signal in state $\theta$. The probability that every
agent abstains in that state is $\prod_{t\geq1}[1-b_\theta h(O_t)]$. Every factor
is positive, and $\sum_t b_\theta h(O_t)<\infty$, so the product is positive as
well. On this event no one tests, yet at each finite date another abstention
conveys information because the next agent could still choose to test.

\clearpage
\begin{center}
{\Large Online Appendix B to \emph{Verification and the Risk of a Wrong Cascade}\par}
\end{center}
\begin{abstract}
We calculate the terminal probabilities in the paper. At signal precision $3/4$,
histories still active at a finite date bound the error in the reported
probabilities. For the comparisons involving the bounty and the gap in testing, we
give the public records at every reachable belief and derive the exit probabilities
from them. The accompanying programs use rational arithmetic.
\end{abstract}

\setcounter{table}{0}
\section[Online Appendix B. Terminal probability calculations]{Terminal probability calculations}\label{app:calculations}

\subsection{From an observed record to an exit probability}
A private signal agrees with the state $\theta\in\{0,1\}$ with probability $q$. A
test succeeds with probability $0<p<1$ in state one and always fails in state zero.
At public odds $O=\mu/(1-\mu)$ on state one, a low signal gives private odds
$O/\Lambda$ and a high signal gives $\Lambda O$, where $\Lambda=q/(1-q)$. An agent
at belief $x$ investigates precisely when
\[
 xp(1+V)+\max\{x(1-p),1-x\}-c>\max\{x,1-x\}.
\]
Use $R(\mu)$ for $V$ if the payment depends on public belief. After a failure the
agent takes action one exactly when $(1-p)O(x)\geq1$, where $O(x)=x/(1-x)$.

Write $N_a$ for abstention and action $a$, and $F_a$ for a test followed by failure
and action $a$. Success establishes state one. At the current odds, let $S(r)$ be
the signals consistent with a feasible record $r\in\{N_0,N_1,F_0,F_1\}$. Put
$b_{\theta s}=\Pr(s\mid\theta)$ and let $f(r)$ equal one for a failure and zero for
abstention. The record probabilities and the new odds are
\begin{align}
 P_0(r)&=\sum_{s\in S(r)}b_{0s},\qquad
 P_1(r)=(1-p)^{f(r)}\sum_{s\in S(r)}b_{1s},\label{eq:recordprob}\\
 O'&=O\,P_1(r)/P_0(r).\label{eq:odds}
\end{align}
If both types investigate and choose the same action after failure, the observer
learns only that the test failed and the odds fall to $(1-p)O$. If the failed
record reveals a favorable signal, the odds become $(1-p)\Lambda O$. 

The public reaches proof ($D$), a permanent cascade on zero ($L$), or a permanent
cascade on one ($U$) at a finite date almost surely in either state (Theorem~\ref{thm:terminal-outcomes} of
the paper). Consider a history $h$ that first enters an interior cascade.
Multiplying \eqref{eq:odds} along $h$ gives its state-one probability as its
state-zero probability times $O(h)/O(\pi)$, the ratio of entry odds to prior odds.

\subsection{The two errors at signal precision three quarters}
Take $q=3/4$, $p=\pi=1/2$, $V=2$, and $c=7/10$ as in Table~\ref{tab:leading-errors} of the paper. With
actions alone the transient odds are $1/3$ and one. Let $b=\Pr_\theta(s=1)$ and let
$u(O)$ be the probability of an upper cascade from odds $O$. A high signal at $1/3$
returns the process to one. A high signal at one starts the upper cascade. Thus
$u(1/3)=b\,u(1)$ and $u(1)=b+(1-b)u(1/3)$, giving $u(1)=12/13$ in state one and
$4/13$ in state zero.

With verification the agent tests at private odds in $(7/8,3/2)\cup(7/3,\infty)$.
Start with unit mass at prior odds one in each state. At each transient belief use
\eqref{eq:recordprob} for the records without proof. In state one the share
$p\sum_{s:\,s\ {\rm tests}}b_{1s}$ goes directly to proof. Set aside the mass
reaching each exit and combine the remaining histories at identical odds. After $n$
agents let $a_{\theta,n}(E)$ be the mass at exit $E$ and $r_{\theta,n}$ the
unresolved mass. Then
\[
 a_{\theta,n}(E)\leq\Pr_\theta(E)
 \leq a_{\theta,n}(E)+r_{\theta,n}.
\]
Before proof all public odds have the form $3^k2^{-j}$ for an integer $k$ and a
nonnegative integer $j$. Table~\ref{tab:bounds} rounds the exact lower bounds down
and upper bounds up.

\begin{table}[htbp]\centering\small
\caption{Bounds on the probability of a wrong lower cascade in state one}
\label{tab:bounds}
\begin{tabular}{@{}rccc@{}}\toprule
Agents & Lower bound & Upper bound & $\max_\theta r_{\theta,n}$ below\\\midrule
10  & 0.140445709228 & 0.148297809065 & $8\times10^{-3}$\\
25  & 0.141488615565 & 0.141492029420 & $3.5\times10^{-6}$\\
50  & 0.141489033704 & 0.141489033715 & $1.1\times10^{-11}$\\
100 & 0.141489033705 & 0.141489033706 & $8\times10^{-23}$\\\bottomrule
\end{tabular}
\end{table}
With no upper trap, $\Pr_1(D)=1-\Pr_1(L)$. At a finite horizon the unresolved mass
in state one may contribute to either exit. In state zero the lower cascade is
reached almost surely. Proposition~\ref{prop:submitted-error-comparison} proves the signs of both comparisons with
actions alone.
\FloatBarrier

\subsection{Fixed and declining payments for proof}
Let $p=\pi=1/2$, $q=2/3$, and $c=1/5$. The fixed bounty $V=1/2$ and the payment
$R(\mu)=1-\mu$, determined before the agent decides to test, coincide at the prior.
Starting at odds one, Table~\ref{tab:reward-graph} gives every reachable belief
before proof. Each cell shows the record and next odds in the absence of success. A
signal inducing a test also yields a success branch with probability $1/2$
conditional on that signal in state one.

\begin{table}[htbp]\centering\small
\caption{All transitions without proof under the two payments}
\label{tab:reward-graph}
\begin{tabular}{@{}lccc@{}}\toprule
Payment & Current odds & Low signal & High signal\\\midrule
Either & $1/8$ & $N_0\to1/8$ & $N_0\to1/8$\\
Either & $1/4$ & $N_0\to1/8$ & $F_0\to1/4$\\
Either & $1/2$ & $N_0\to1/4$ & $F_0\to1/2$\\
Either & $1$   & $F_0\to1/4$ & $N_1\to2$\\
Either & $2$   & $F_0\to1/2$ & $N_1\to4$\\
Fixed & $4$    & $N_1\to2$   & $F_1\to4$\\
Declining & $4$ & $N_1\to4$ & $N_1\to4$\\\bottomrule
\end{tabular}
\end{table}

With the fixed bounty, state zero reaches odds $1/8$ almost surely. After two
initial high signals it is possible to stay at odds four by drawing high signals
forever, an event of probability zero. Since the prior odds are one, the
probability of reaching $1/8$ in state one is $1/8$, leaving $7/8$ for proof. With
the declining payment, two initial highs are necessary and sufficient to enter the
upper cascade at odds four. They have probability $1/9$ in state zero and $4/9$ in
state one. The lower cascade accounts for the remaining $8/9$ in state zero and
hence for $(8/9)(1/8)=1/9$ in state one. This leaves $4/9$ for proof. At these
parameters the payment $R(\mu)=(1-\mu)/2$ gives all the same records as $1-\mu$.

Why does the fixed bounty make the wrong lower cascade more likely in state one?
The policies agree until odds first reach four. From there, continued testing under
the fixed bounty leads to $1/8$ almost surely in state zero. In state one the
conditional probability of this exit is $(1/8)/4=1/32$, leaving $31/32$ for proof.
From the prior the probability of reaching four in state one is $4/9$. The increase
in error is therefore $(4/9)(1/32)=1/72$.
\FloatBarrier

\subsection{Filling the gap while fixing the lower testing cutoff}
Let $p=1/2$, $q=4/7$, and $\pi=2/3$, so the prior odds are two. At
$(V,c)=(1/2,1/5)$ the agent tests at private beliefs in $(4/15,16/25)\cup(4/5,1]$.
At $(V,c)=(1,4/15)$ she tests throughout $(4/15,1]$. Adjusting the cost as we raise
the bounty fixes the lower cutoff. Only beliefs in $[16/25,4/5]$ change the
decision.

\begin{table}[htbp]\centering\small
\caption{Transitions from prior odds two under the two testing rules}
\label{tab:gap-graph}
\begin{tabular}{@{}lccc@{}}\toprule
Testing rule & Current odds & Low signal & High signal\\\midrule
Gap & $2$ & $F_0\to3/4$ & $N_1\to8/3$\\
Gap filled & $2$ & $F_0\to3/4$ & $F_1\to4/3$\\\midrule
Either & $3/4$ & $F_0\to3/8$ & $F_0\to3/8$\\
Either & $3/8$ & $N_0\to9/32$ & $F_0\to1/4$\\
Either & $9/32$ & $N_0\to27/128$ & $F_0\to3/16$\\\midrule
Gap filled & $4/3$ & $F_0\to2/3$ & $F_0\to2/3$\\
Gap filled & $2/3$ & $F_0\to1/3$ & $F_0\to1/3$\\
Gap filled & $1/3$ & $N_0\to1/4$ & $F_0\to2/9$\\\bottomrule
\end{tabular}
\par\smallskip\begin{minipage}{.96\linewidth}\footnotesize
The remaining odds are absorbing. The upper exit is $8/3$ under the
original rule. The lower exits are $1/4$, $27/128$, and $3/16$ under
both rules, with $2/9$ an additional lower exit under the filled rule.
\end{minipage}
\end{table}

With the gap in place, a first high signal is necessary and sufficient for the
upper cascade. In state zero the three lower entries $1/4$, $27/128$, and $3/16$
have respective probabilities $12/49$, $64/343$, and $48/343$. Weighting each by
its entry odds and dividing by the prior odds of two gives $87/1372$ for the
state-one lower cascade. Filling the gap leaves the initial low branch as before
but leads the first high type to test. Conditional on that high in state one, a
lower cascade now requires three failures, followed either by a low signal or by a
high and a fourth failure. Its probability on this branch is
$(1/2)^3[(3/7)+(4/7)(1/2)]=5/56$. The initial high has probability $4/7$, so
filling the gap adds $5/98$ to the state-one error and $51/98$ to proof. The last
row of Table~\ref{tab:gap-exits} gives the comparison when the cost is held fixed
and the lower cutoff changes.

\begin{table}[htbp]\centering\small
\caption{Terminal probabilities when the gap is filled}
\label{tab:gap-exits}
\begin{tabular}{@{}lrrrr@{}}\toprule
 &\multicolumn{3}{c}{State 1}&State 0\\
\cmidrule(lr){2-4}\cmidrule(l){5-5}
$(V,c)$ &Proof &Lower &Upper &Upper\\\midrule
$(1/2,1/5)$ & $501/1372$ & $87/1372$ & $4/7$ & $3/7$\\
$(1,4/15)$, same cutoff & $1215/1372$ & $157/1372$ & 0 & 0\\
$(1,1/5)$, same cost & $2497/2744$ & $247/2744$ & 0 & 0\\\bottomrule
\end{tabular}
\end{table}

For this last row, set $V=1$ and keep $c=1/5$. In state zero an initial low takes
the odds along $2,3/4,3/8,3/16$ to a lower cascade. After an initial high the
corresponding path is $2,4/3,2/3,1/3$. The next two signals determine the exit from
$1/3$. Two lows yield odds $3/16$, one of each gives $1/6$, and two highs give
$4/27$. Conditional on reaching $1/3$ in state zero, the expected odds at entry are
\[
 (4/7)^2(3/16)+2(4/7)(3/7)(1/6)+(3/7)^2(4/27)=25/147 .
\]
Weighting the two initial branches in state zero and dividing by the prior odds now
gives $\Pr_1(L)=[(4/7)(3/16)+(3/7)(25/147)]/2=247/2744$.
\FloatBarrier

\begin{thebibliography}{99}

\bibitem[Ali(2018)]{Ali2018}
Ali, S.N. (2018).
Herding with costly information.
\textit{Journal of Economic Theory} 175, 713--729.
\url{https://doi.org/10.1016/j.jet.2018.02.009}.

\bibitem[Banerjee(1992)]{Banerjee1992}
Banerjee, A.V. (1992).
A simple model of herd behavior.
\textit{Quarterly Journal of Economics} 107(3), 797--817.
\url{https://doi.org/10.2307/2118364}.

\bibitem[Bikhchandani et~al.(1992)]{BikhchandaniHirshleiferWelch1992}
Bikhchandani, S., Hirshleifer, D., Welch, I. (1992).
A theory of fads, fashion, custom, and cultural change as informational
cascades.
\textit{Journal of Political Economy} 100(5), 992--1026.
\url{https://doi.org/10.1086/261849}.

\bibitem[Bikhchandani et~al.(2024)]{BikhchandaniHirshleiferTamuzWelch2024}
Bikhchandani, S., Hirshleifer, D., Tamuz, O., Welch, I. (2024).
Information cascades and social learning.
\textit{Journal of Economic Literature} 62(3), 1040--1093.
\url{https://doi.org/10.1257/jel.20241472}.

\bibitem[Burguet and Vives(2000)]{BurguetVives2000}
Burguet, R., Vives, X. (2000).
Social learning and costly information acquisition.
\textit{Economic Theory} 15(1), 185--205.
\url{https://doi.org/10.1007/s001990050006}.

\bibitem[Frazier et~al.(2014)]{FrazierKempeKleinbergKleinberg2014}
Frazier, P., Kempe, D., Kleinberg, J., Kleinberg, R. (2014).
Incentivizing exploration.
In \textit{Proceedings of the Fifteenth ACM Conference on Economics and
Computation}, 5--22.
\url{https://doi.org/10.1145/2600057.2602897}.

\bibitem[Halac et~al.(2017)]{HalacKartikLiu2017}
Halac, M., Kartik, N., Liu, Q. (2017).
Contests for experimentation.
\textit{Journal of Political Economy} 125(5), 1523--1569.
\url{https://doi.org/10.1086/693040}.

\bibitem[Hendricks et~al.(2012)]{HendricksSorensenWiseman2012}
Hendricks, K., Sorensen, A., Wiseman, T. (2012).
Observational learning and demand for search goods.
\textit{American Economic Journal: Microeconomics} 4(1), 1--31.
\url{https://doi.org/10.1257/mic.4.1.1}.

\bibitem[Ivanik and Lukyanov(2025)]{IvanikLukyanov2025}
Ivanik, V., Lukyanov, G. (2025).
Contrarian incentives and costly social learning.
arXiv:2508.21446, version 9, October 2026.
\url{https://arxiv.org/abs/2508.21446v9}.

\bibitem[Keller et~al.(2005)]{KellerRadyCripps2005}
Keller, G., Rady, S., Cripps, M. (2005).
Strategic experimentation with exponential bandits.
\textit{Econometrica} 73(1), 39--68.
\url{https://doi.org/10.1111/j.1468-0262.2005.00564.x}.

\bibitem[Mueller-Frank and Pai(2016)]{MuellerFrankPai2016}
Mueller-Frank, M., Pai, M.M. (2016).
Social learning with costly search.
\textit{American Economic Journal: Microeconomics} 8(1), 83--109.
\url{https://doi.org/10.1257/mic.20130253}.

\bibitem[Smith and S{\o}rensen(2000)]{SmithSorensen2000}
Smith, L., S{\o}rensen, P. (2000).
Pathological outcomes of observational learning.
\textit{Econometrica} 68(2), 371--398.
\url{https://doi.org/10.1111/1468-0262.00113}.

\bibitem[Smith et~al.(2021)]{SmithSorensenTian2021}
Smith, L., S{\o}rensen, P.N., Tian, J. (2021).
Informational herding, optimal experimentation, and contrarianism.
\textit{Review of Economic Studies} 88(5), 2527--2554.
\url{https://doi.org/10.1093/restud/rdab001}.

\end{thebibliography}
\end{document}